\documentclass[11pt,preprint,preprintnumbers,a4paper]{article}
\pdfoutput=1
             
\usepackage{jheppub}

\usepackage{graphicx,array}
\usepackage{color}
\usepackage{latexsym}
\usepackage{amsthm}
\usepackage{amsmath}
\usepackage{amssymb}
\usepackage{hyperref}
\usepackage{bbold}
\usepackage{subfig}
\usepackage{multirow}
\usepackage{slashed}
\usepackage[normalem]{ulem}
\usepackage[makeroom]{cancel}
\usepackage{hyperref}

\font\tenrsfs=rsfs10 at 12pt

\font\sevenrsfs=rsfs7
\font\fiversfs=rsfs5
\newfam\rsfsfam
\textfont\rsfsfam=\tenrsfs
\scriptfont\rsfsfam=\sevenrsfs
\scriptscriptfont\rsfsfam=\fiversfs
\def\mathscr#1{{\fam\rsfsfam\relax#1}}

\font\tenbbm=bbm10 at 12pt
\newfam\bbmfam
\textfont\bbmfam=\tenbbm

\definecolor{darkblue}{cmyk}{1,0.3,0,0.2}
\definecolor{violet}{cmyk}{0,1,0,0.2}
\hypersetup{colorlinks, bookmarksnumbered, citecolor=darkblue, linkcolor=darkblue, pdfstartview=FitH, urlcolor=darkblue, linktocpage}

\newcommand{\TeV}{{\rm TeV}}
\newcommand{\SM}{{\rm SM}}
\newcommand{\SU}{{\rm SU}}

\newcommand{\arXhref}[1]{\href{http://arxiv.org/abs/#1}{{\tt #1}}}

\newcommand{\fref}[1]{Fig.~\ref{fig:#1}} 
\newcommand{\eref}[1]{Eq.~\eqref{eq:#1}}

\newcommand{\cref}[1]{Chapter~\ref{ch:.#1}}

\newcommand{\beq}{\begin{equation}} 
\newcommand{\eeq}{\end{equation}} 
\newcommand{\ba}{\begin{array}}  
\newcommand{\ea}{\end{array}} 
\newcommand{\bea}{\begin{eqnarray}}  
\newcommand{\eea}{\end{eqnarray}}  
\newcommand{\be}{\begin{equation}}  
\newcommand{\ee}{\end{equation}}  
\newcommand{\bal}{\begin{align}}
\newcommand{\eal}{\end{align}}   
\newcommand{\bi}{\begin{itemize}}  
\newcommand{\ei}{\end{itemize}}  
\newcommand{\ben}{\begin{enumerate}}  
\newcommand{\een}{\end{enumerate}}  
\newcommand{\bc}{\begin{center}}
\newcommand{\ec}{\end{center}} 
\newcommand{\bt}{\begin{table}}
\newcommand{\et}{\end{table}}  
\newcommand{\btb}{\begin{tabular}}
\newcommand{\etb}{\end{tabular}}  
\newcommand{\bvec}{\left ( \ba{c}}
\newcommand{\evec}{\ea \right )}

\newcommand{\cO}{{\mathcal O}} 
 
\newcommand{\cL}{{\mathcal L}}

\newcommand{\cA}{{\mathcal A}}

\newcommand{\fb}{\mathrm{fb}}

\newcommand{\ds}{\displaystyle}

\newcommand{\lz}{\lambda_z}
\newcommand{\dgz}{\delta g_{1,z}}
\newcommand{\dkg}{\delta \kappa_\gamma}

\newcommand{\adam}[1]{{\color{black} #1}}
\newcommand{\admir}[1]{{\color{black} #1}}
\newcommand{\david}[1]{{\color{black} #1}}
\newcommand{\martin}[1]{{\color{black} #1}}

\preprint{ZU-TH-34/16}

\title{
Anomalous Triple Gauge Couplings in the Effective Field Theory Approach at the LHC
}

\author[a]{Adam Falkowski,}
\author[b]{Mart\'{i}n Gonz\'{a}lez-Alonso,}
\author[c,d]{Admir Greljo,}
\author[c]{David Marzocca,}
\author[e]{and Minho Son}

\affiliation[a]{Laboratoire de Physique Th\'{e}orique, Bat.~210, Universit\'{e} Paris-Sud, 91405 Orsay, France}
\affiliation[b]{IPN de Lyon/CNRS, Universite Lyon 1, Villeurbanne, France}
\affiliation[c]{Physik-Institut, Universit\"at Z\"urich, CH-8057 Z\"urich, Switzerland}
\affiliation[d]{Faculty of Science, University of Sarajevo, Zmaja od Bosne 33-35, \\ 71000 Sarajevo, Bosnia and Herzegovina}
\affiliation[e]{Department of Physics, Korea Advanced Institute of Science and Technology,\\291 Daehak-ro, Yuseong-gu, Daejeon 34141, Republic of Korea}

\emailAdd{adam.falkowski@th.u-psud.fr}
\emailAdd{m.gonzalez@ipnl.in2p3.fr}
\emailAdd{admir@physik.uzh.ch}
\emailAdd{marzocca@physik.uzh.ch}
\emailAdd{minho.son@kaist.ac.kr}

\abstract{
We discuss how to perform consistent extractions of anomalous triple gauge couplings (aTGC) from electroweak boson pair production at the LHC in the Standard Model Effective Field Theory (SMEFT). After recasting recent ATLAS and CMS searches in $pp\to WZ (WW)\to \ell' \nu \ell^+\ell^- (\nu_{\ell})$ channels, we find that: (a) working consistently at order $\Lambda^{-2}$ in the SMEFT expansion the existing aTGC bounds from Higgs and LEP-2 data are not improved, (b) the strong limits quoted by the experimental collaborations are due to the partial $\Lambda^{-4}$ corrections (dimension-6 squared contributions). Using helicity selection rule arguments we are able to explain the suppression in some of the interference terms, and discuss conditions on New Physics (NP) models that can benefit from such LHC analyses. Furthermore, standard analyses assume implicitly a quite large NP scale, an assumption that can be relaxed by imposing cuts on the underlying scale of the process ($\sqrt{\hat{s}}$). In practice, we find almost no correlation between $\sqrt{\hat{s}}$ and the experimentally accessible quantities, which complicates the SMEFT interpretation. Nevertheless, we provide a method to set (conservative) aTGC bounds in this situation, and recast the present searches accordingly. Finally, we introduce a simple NP model for aTGC to compare the bounds obtained directly in the model with those from the SMEFT analysis.

}

\begin{document}

\maketitle
\flushbottom

\section{Introduction}
\label{sec:intro}

Cubic and quartic self-interactions of the electroweak gauge bosons are present in the Standard Model (SM) due to the underlying non-abelian gauge symmetry, and are completely fixed by the gauge couplings, namely, the electromagnetic coupling constant $e$ and the weak mixing angle $s_\theta \equiv \sin \theta_W$. This, however, is not the case in a general Beyond the Standard Model (BSM) scenario. Therefore, processes that are sensitive to gauge boson self-interactions are important tools used to search for nonstandard effects.

In this work we focus on general BSM contributions to the cubic electroweak gauge bosons interactions, employing the linear Effective Field Theory (EFT) framework, also known as the Standard Model Effective Field Theory (SMEFT). In this model-independent approach, the SM (with the Higgs embedded in an $\SU(2)_L$ doublet) is extended by non-renormalizable gauge-invariant operators with canonical dimensions $D > 4$ which encode the effects of some new physics with a mass scale $\Lambda$ much larger than the electroweak scale. The BSM effects are thus organized as an expansion in $1/\Lambda$, and the leading lepton-number-conserving terms are $\cO(\Lambda^{-2})$ generated by $D=6$ operators in the SMEFT Lagrangian:
\be
	\cL^{\rm eff} = \cL_{\rm SM} + \sum_i \frac{c_i^{(6)}}{\Lambda^2} \cO_i^{(6)} + \sum_j \frac{c_j^{(8)}}{\Lambda^4} \cO_j^{(8)} + \ldots ~.
	\label{eq:eff_lagr_general}
\ee

We are interested in diboson production at the LHC, which in general is sensitive to many (linear combinations of) effective operators. They can affect the process through their modifications of the couplings of gauge bosons to fermions, the gauge boson propagators or the cubic interactions of the gauge bosons. However, once we take into account LEP1 constraints~\cite{Falkowski:2014tna,Efrati:2015eaa}, CP-conserving observables in diboson production are effectively controlled by 3 combinations of  EFT parameters  at ${\cal O}(\Lambda^{-2})$ in the SMEFT, which we can choose to be the 3 anomalous Triple Gauge Couplings (aTGC), $\left\{ \delta g_{1,z},\delta \kappa_\gamma, \lambda_z \right\}$, defined as follows~\cite{DeRujula:1991ufe,Hagiwara:1993ck}:
\bea  
\label{eq:atgc}
 \cL_{\rm tgc}  &=& 
i  e    \left ( W_{\mu \nu}^+ W_\mu^-  -  W_{\mu \nu}^- W_\mu^+ \right ) A_\nu    
  +  i  e {c_\theta \over s_\theta} 
 \left (1 + \delta g_{1,z} \right )   \left ( W_{\mu \nu}^+ W_\mu^-  -  W_{\mu \nu}^- W_\mu^+ \right ) Z_\nu \nonumber \\
&+& i e (1 + \delta \kappa_\gamma)  A_{\mu\nu}\,W_\mu^+W_\nu^-   +   i  e {c_\theta \over s_\theta}  \left (1 +  \delta \kappa_z \right ) Z_{\mu\nu}\,W_\mu^+W_\nu^-  \nonumber \\   &+&     
 i   { \lambda_z  e  \over m_W^2 } \left [   W_{\mu \nu}^+W_{\nu \rho}^- A_{\rho \mu}  +  {c_\theta \over s_\theta} W_{\mu \nu}^+W_{\nu \rho}^- Z_{\rho \mu}   \right]~,
\eea 
where $c_\theta = \sqrt{1 - s_\theta^2}$~, $\delta \kappa_z  = \delta g_{1,z} -{s_\theta^2 \over c_\theta^2} \delta \kappa_\gamma$. These aTGC can be computed in function of Wilson coefficients of $D = 6$ operators in \eref{eff_lagr_general}, and they are formally of order \footnote{\david{See App.~\ref{app:HelAmp} for the explicit dependence of the aTGC in Eq.~(\ref{eq:atgc}) on the gauge-invariant operators in the Warsaw \cite{Grzadkowski:2010es} and SILH \cite{Giudice:2007fh} bases.}}
\be
	\delta g_{1,z},~ \delta \kappa_\gamma,~ \lambda_z~ \sim c^{(6)} \frac{m_W^2}{\Lambda^2}~,
\ee
so that in the SM limit all three aTGC vanish. Let us stress that in deriving this matching one should be careful to redefine fields and input parameters in a way which satisfies the property that after imposing LEP-1 bounds the aTGC are the only three unconstrained parameters relevant to diboson production (see e.g. Refs.~\cite{Pomarol:2013zra,Gupta:2014rxa,Falkowski:2014tna,HiggsBasis,Trott:2014dma,deFlorian:2016spz}). The dictionary between the aTGCs and Wilson coefficients of $D=6$ operators in various bases can be found in Appendix~\ref{app:HelAmp} (from Ref.~\cite{HiggsBasis}).

Any experimental observable (such as differential cross section, number of signal events in a bin, etc.) obtained from the effective Lagrangian in Eq.~\eqref{eq:eff_lagr_general} takes the following form
\be
	\sigma = \sigma^{\SM} + \sum_i \left(\frac{c^{(6)}_i }{\Lambda^2} \sigma_i^{\rm  (6\times SM)} + \rm{h.c.} \right) + \sum_{ij} \frac{c^{(6)}_i c^{(6)*}_j }{\Lambda^4} \sigma_{ij}^{(6\times 6)}  + \sum_j \left(\frac{c^{(8)}_j }{\Lambda^4} \sigma_j^{\rm (8\times SM)} + \rm{h.c.} \right) + \ldots~.
\ee
It is important to notice that the $D=6$ squared terms are of the same order in the EFT expansion parameter $\Lambda$ as the (neglected) interference of the $D=8$ with the SM.

Precision constraints on aTGCs can be derived from $W^+W^-$ production in LEP-2~\cite{Schael:2013ita}, see e.g. \cite{Falkowski:2014tna,Berthier:2016tkq} for EFT interpretations.
Meanwhile,  it has been pointed out that the LHC Higgs data can also lead to meaningful indirect constraints on the aTGC in the context of SMEFT~\cite{Corbett:2013pja,Pomarol:2013zra,Dumont:2013wma,Masso:2014xra,Ellis:2014jta,Ellis:2014dva}. \martin{This becomes evident when the effective operators that generate the aTGC defined in Eq.~\eqref{eq:atgc} are written in an explicitly gauge-invariant form, since they involve not only gauge bosons but also the $SU(2)_L$ Higgs doublet (see Eq.~\ref{eq:warsaw_atgc} and \ref{eq:PEL_tgc}).} Recently, Ref.~\cite{Falkowski:2015jaa} reported a global fit in the SMEFT to LEP-2 $WW$ and LHC Higgs signal-strength data, by working consistently at $\cO(\Lambda^{-2})$. In particular, the analysis considered only $D=6$ operator interference with the SM, under the Minimal Flavor Violation (MFV) assumption, in which case the full set of relevant  linear combinations of $D=6$ operator affecting that analysis is limited to ten.
The result of that fit projected to aTGC is
\beq\begin{split}
\label{eq:constraints}
\bvec 
\delta g_{1,z} \\ \delta \kappa_\gamma \\ \lambda_z 
\evec   &= \bvec 0.043  \pm  0.031 \\ 0.142 \pm 0.085 \\  -0.162 \pm 0.073 \evec, \;
\rho = \left ( \ba{ccc} 
1 & 0.74 & -0.85 \\ 
0.74 & 1 & -0.88 \\
-0.85 & -0.88 & 1 
\ea \right ).
\end{split}
\eeq
Interestingly enough, the combination of the two datasets lifts the flat direction present in each of them taken separately~\cite{Falkowski:2015jaa}. As a result, the bounds do not change significantly when the (formally subleading) dim-6 squared contributions are included in the analysis. Thus, these results constitute robust and model-independent bounds on the aTGC. They can be easily translated to any given BSM model (that can be matched to the SMEFT) to set bounds on the corresponding masses and couplings without having to re-do the analysis of the data. 

It is well-known that $W^+W^-$ and $W^\pm Z$ differential production cross sections at Tevatron and LHC are also very sensitive to aTGC~\cite{Abazov:2012ze,Aaltonen:2012vu,Khachatryan:2015sga,Aad:2016wpd}. In addition, recent progress on NNLO QCD predictions in the SM~\cite{Grazzini:2016ctr,Grazzini:2016swo} facilitate the study of BSM effects. 
However, these measurements were not included in the previous global analysis of Ref.~\cite{Falkowski:2015jaa}, because their EFT interpretation is much more involved. 
\adam{One technical issue was that the combination with prior LHC bounds on aTGC was not possible because these were not performed with all three anomalous couplings present simultaneously and/or the associated likelihood was not provided (i.e. the correlation matrix if the distribution is gaussian).\footnote{That issue was properly addressed in more recent Ref.~\cite{Butter:2016cvz}.} }
But the main complication comes from the fact that hadron collisions probe a wide range of energies. 
This is in contrast with LEP-2 observables and on-shell Higgs decay measurements, where the typical energy scale is bounded by the LEP center-of-mass energy and Higgs mass, respectively.
\adam{In the LHC case, the EFT expansion is more slowly convergent because $\hat{s}/\Lambda^2$ can be large toward the tail of differential distributions.}
This enhances the sensitivity to neglected dim-8 operators and complicates the extraction of robust aTGC bounds.
\adam{For this reason, the question of the {\em validity regime of the EFT approach} have to be carefully addressed to properly interpret  aTGC constraints extracted from  $W^+W^-$ and $W^\pm Z$ measurements in hadron colliders.}
 
\adam{
Let us clarify here what we understand by the EFT validity regime.   
The relevant question here is whether the constraints on the aTGCs can be translated into constraints on masses and couplings of new particles in extensions of the SM.   
By construction, the EFT provides a good approximation of the underlying UV theory at energy scales $E \ll \Lambda$. However, from low energy measurements one can only extract the combination $c/\Lambda^2$, where $c$ is the Wilson coefficient of the relevant operator. 
Therefore the discussion of the validity for a given experimental energy $E$ requires assumptions on the magnitude of $c$, and is thus necessarily model dependent. 
At the end of the day, given the energy scale and precision of the experiment, the validity discussion amounts to formulating a set of conditions under which the EFT results can be used to constrain BSM models.}  

\adam{
One test of validity is to compare the magnitude of linear and quadratic contributions of $D$=6 operators to observables. 
The dimension-6 squared contributions are formally $\cO(\Lambda^{-4})$ in the EFT expansions, and thus they are expected to be of the same order as the linear contributions of the neglected $D$=8 operators. 
If the linear $D$=6 contributions dominate, which is the case for the analysis of Ref.~\cite{Falkowski:2015jaa}, then the EFT results are robust and can be used to constrain any BSM model satisfying the minimal EFT assumptions, namely a linear EWSB and $\Lambda \gg E$.  
Last but not least, the small sensitivity to dim-6 squared contributions ensures that the results are basis-independent, as different bases of $D=6$ operators in the literature differ by $\cO(\Lambda^{-4})$ terms.
Conversely, if the squared contributions were important, these results would not constitute valid bounds in the most general case, and a consistent EFT interpretation of the data would require some more assumptions about the UV models. 
}

\adam{
It turns out that  the bounds on aTGCs obtained from the LHC diboson measurements strong rely on the inclusion of $\cO(\Lambda^{-4})$ dim-6 squared contributions \cite{Butter:2016cvz}. 
The situation is further worsened because the linear effects of dim-6 operators (coming from its interference with the SM) happen to be suppressed in these observables (for a general discussion see Ref.~\cite{Azatov:2016sqh} and for the particular observables used here see Section~\ref{subsec:noninterference}).
In this context it is important to stress that the small sensitivity to quadratic terms is  not a necessary condition to ensure the EFT validity\martin{, i.e. its applicability to certain BSM scenarios. In fact,}} 
as discussed in Refs.~\cite{Biekoetter:2014jwa,Contino:2016jqw}\martin{,} in a wide class of BSM models with some strongly coupled sector the contribution from dim-8 operators is subleading with respect to dim-6 squared terms without invalidating the EFT expansion.
This can be understood from a simple matching of the Wilson coefficients to the UV parameters of the theory:  $c_i^{(6)} \sim c_j^{(8)} \sim g_*^2$,  
\adam{
where $g_*$ denotes the coupling strength of the SM currents to the BSM resonances.}\footnote{
\adam{See Section~4 for a particular example. 
More generally, given some broad assumptions about the UV theory, one can deduce the dependence of the EFT Wilson coefficients on the couplings strength $g_*$ characterizing the strongly interacting sector  \cite{Giudice:2007fh,Liu:2016idz}. 
}}
  
This implies that,  if $g_*\gg 1$,  the dim-6 squared terms dominate over the linear dim-8 by a factor $g_*^2 / g_{\rm SM}^2 \gg 1$.
Consequently, ``standard" aTGC analysis of LHC data is justified for these BSM scenarios. Even in such cases it is convenient to perform the EFT analysis using different cuts on the appropriate kinematical variables~\cite{Biekoetter:2014jwa,Contino:2016jqw}. In this way,  the applicability of the EFT analysis is extended to a wider range of BSM models in which a new state is not far from the scales being probed at the LHC. 
However, the relevant variable that controls the validity range of EFT (partonic center-of-mass energy $\hat{s}$) turns out to be hard to reconstruct experimentally. 
This is evident for the $pp \to WW \to \ell \ell \nu \nu$ process,  where the presence of two neutrinos in the final state impedes unambiguous determination of $\hat s$,  but even for $pp\to WZ \to \ell' \nu \ell^+\ell^-$ where, while reconstructing $\hat s$ is straightforward in theory, experimental uncertainties severely limit the usefulness of such a procedure.
We will evaluate the possibility of using other measurable quantities instead in order to consistently set bounds on aTGC in this situation. 

It is the purpose of this work to discuss these issues in some detail, and study what are their implications for the aTGC bounds obtained from LHC data.
In particular, in Section \ref{sec:EFTLimitSetting} we discuss the importance of dim-6 squared terms in diboson production, how to derive bounds consistently within the EFT when the center-of-mass energy of the process is not directly observable, and provide an analysis of the interference between SM and BSM amplitudes. In Section \ref{sec:recast} we use these methods to recast a selection of ATLAS and CMS $WW$ and $WZ$ analysis, both with 8~TeV and 13~TeV data, in order to extract consistent bounds on aTGC. In Section \ref{sec:model} we provide an explicit example of a BSM model generating aTGC, in order to compare the constraints on the model parameters obtained directly from simulating events using the model with the indirect ones from the aTGC analysis. Finally, we conclude in Section \ref{sec:conclusions}. The two Appendices \ref{app:NonInterferenceSMandBSM} and \ref{app:HelAmp} include a detailed discussion on the helicity amplitudes relevant to diboson production.

\section{Considerations about the EFT validity}
\label{sec:EFTLimitSetting}

\subsection{Total cross section of $WW$ and $WZ$ processes}
\label{subsec:xsecWWandWZ}
Before performing the complex numerical analysis of LHC data, it is convenient to have an initial look at the relevant total cross sections and their naive sensitivity to aTGC. As mentioned in the Introduction, these observables are also sensitive to other nonstandard effects, such as those modifying the $Z$ and $W$ propagators, or their couplings to light fermions. However, given the model-independent constraints from electroweak precision data~\cite{Efrati:2015eaa}, the $WW$ and $WZ$ cross sections effectively constrain 3 linear combinations of Wilson coefficients of dim-6 operators that correspond to the aTGC~\cite{Falkowski:2014tna}. Thus, we have
\be
\sigma  =  \sigma_{SM} \big [ 1 + B_a \kappa_a +  C_{ab}\kappa_a \kappa_b \big ] ~,
\label{eq:tgcdep}
\ee
where $a$ and $b$ run over the three aTGC $\kappa \equiv \left\{ \lz, \dgz, \dkg\right\}$, and $\sigma$ denotes $\sigma(pp\rightarrow W^+W^-)$ or $\sigma(pp\rightarrow W^\pm Z)$.

\begin{figure}[tb]
\centering
\includegraphics[width=0.32\textwidth]{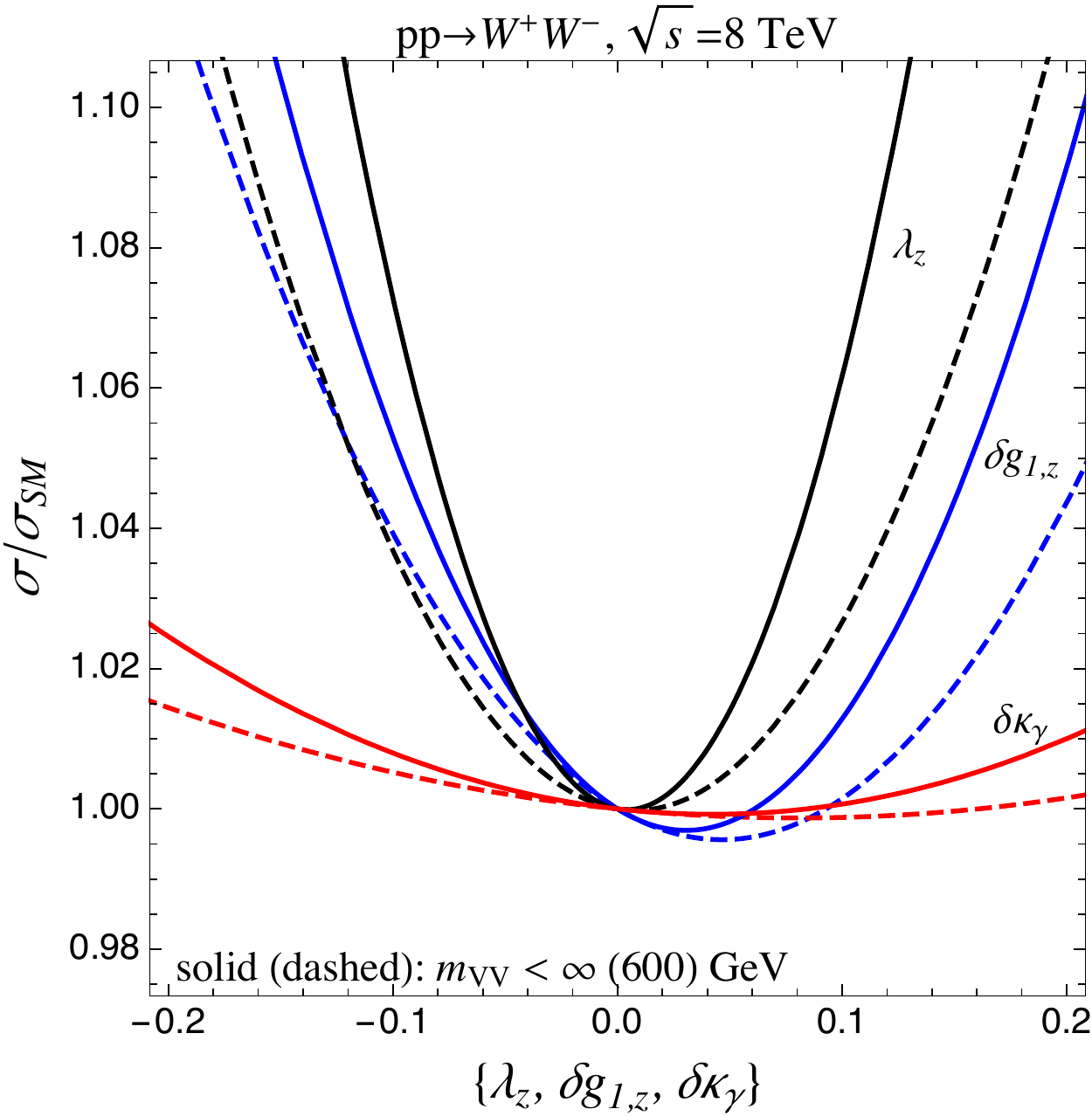}
\includegraphics[width=0.32\textwidth]{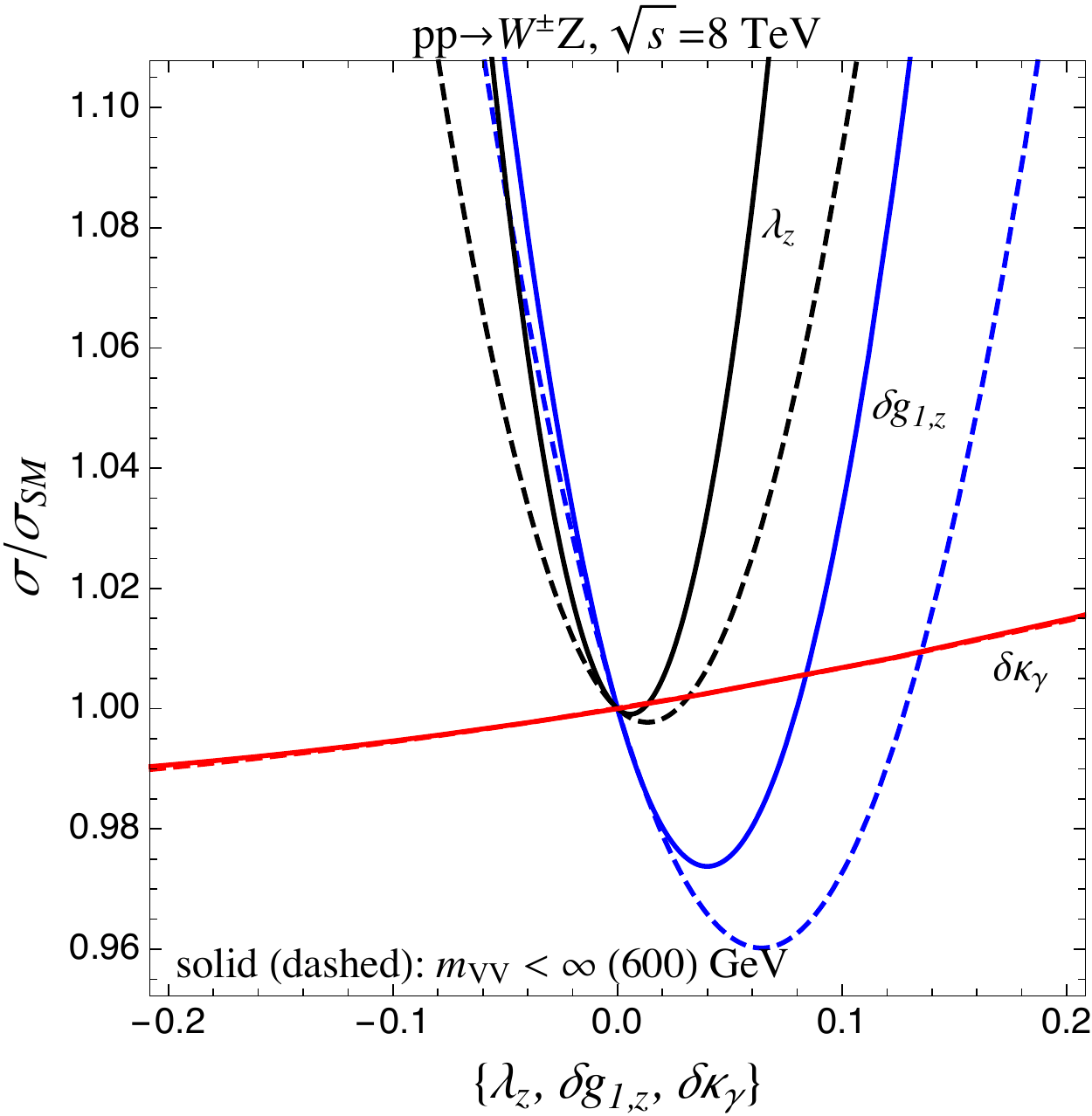}
\includegraphics[width=0.32\textwidth]{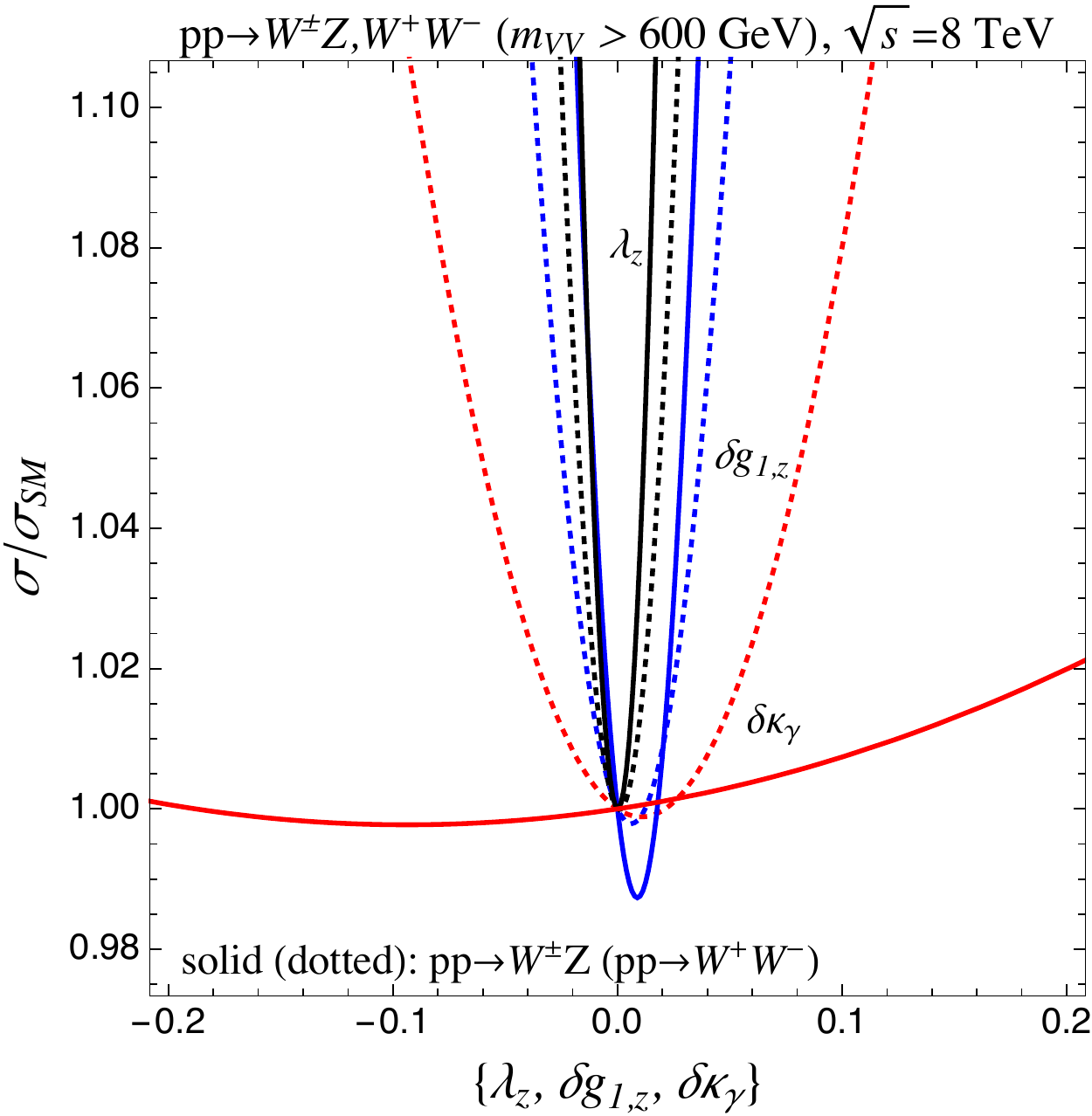}
\caption{Dependence of the $\sigma(pp\rightarrow WW)$ (left) and $\sigma(pp\rightarrow WZ)$ (middle) on the aTGC, $\lambda_z$ (black), $\delta g_{1,z}$ (blue), and $\delta \kappa_\gamma$ (red). One parameter is varied at a time while the other two are set to zero. In the left and center panels the solid (dashed) lines correspond to the cases with $m_{VV} (\equiv\sqrt{\hat{s}}) < \infty$ (600 GeV). In the right panel, instead, only high energy events ($m_{VV} > 600$ GeV) are shown, using solid (dotted) lines for $pp\rightarrow WZ (WW)$.}
\label{fig:xsec_aTGC}
\end{figure}

In Fig.~\ref{fig:xsec_aTGC} we plot the relative cross sections of $WW$ and $WZ$ processes with respect to the SM one at $\sqrt{s} = 8$ TeV  by varying one parameter at a time while keeping the other two at zero. Although the plots are missing the effects from the cross terms between different parameters, they illustrate some important features. First of all, Fig.~\ref{fig:xsec_aTGC} shows that both $WW$ and $WZ$ processes are very sensitive to $\lambda_z$ and $\delta g_{1,z}$, but not so much to $\delta \kappa_\gamma$, which will be rather weakly constrained. We also observe that the $WZ$ channel seems to be more sensitive than the $WW$ one, at least concerning $\lambda_z$ and $\delta g_{1,z}$.

The solid lines, which represent the total cross sections without any cut, show clearly that the quadratic terms in Eq.~\eqref{eq:tgcdep} are not negligible at all. Taking into account that the typical experimental precision in this observable is in the few per-cent ballpark, one can see that the extracted aTGC bounds will be completely dominated by these quadratic effects. As briefly discussed in the Introduction, this is somewhat expected given the high energy scales probed by these processes.
In the case of $\lambda_z$ it is striking to notice that the interference term is almost vanishing. This can be understood by studying the relevant SM and BSM helicity amplitudes, as discussed in Section~\ref{subsec:noninterference}.

In order to analyze the effect of removing the events in the high energy tail, the dashed lines in the left and center panels of Fig.~\ref{fig:xsec_aTGC} show the weakened sensitivity when the cross sections are obtained using only the events with $\sqrt{\hat{s}} < 600$ GeV. Although the effect of the cut is clearly visible, the quadratic effects still remain very important. 
We have also checked that this is still true for a cut as low as 300 GeV. For completeness, in the right panel we show with solid (dotted) lines the $WZ$ ($WW$) cross section for high-energy events ($\sqrt{\hat{s}} > 600$ GeV) only. It is clear that in this region the quadratic terms largely dominate over the linear ones, as expected. The situation is further complicated by the fact that imposing this type of cut on the real data is by no means easy, as we discuss in the next section.

\subsection{Limiting the physical scale of the process }

As already mentioned, the relevant energy scale of diboson production processes is the $VV$ invariant mass, $\sqrt{\hat{s}}\ (\equiv m_{VV})$. 
The differential cross section,
$d\sigma/d m_{VV}$, is therefore a very sensitive probe to new physics effects, and has the potential to disentangle the different aTGC parameters.
A few challenges arise in consistently setting limits on BSM from data.  First, the EFT approach is only valid sufficiently below a cut-off scale corresponding to the mass of new states. Since such scale is not known a priori, various choices of cut-off scales need to be implemented while setting limits within the EFT framework.
Ideally, if the full invariant mass of the $VV$ system (or equivalently $\sqrt{\hat{s}}$) could be reconstructed from data, one would impose an appropriate cut on $m_{VV}$ on both data and simulated events, allowing to build the likelihood using expected and observed cross sections with the cuts, i.e.
\begin{equation}\label{eq:idealLimit}
  (\sigma_{\SM} + \sigma_{\rm BSM})(m_{VV}<m_{VV}^{\rm max})~, \qquad
  \sigma_{\rm obs}(m_{VV}<m_{VV}^{\rm max})~.
\end{equation}
In this way one would derive bounds consistently, with the EFT applicable to theories in which new states are heavier than $m_{VV}^{\rm max}$.
Note that $\sigma_{\rm BSM}$ in Eq.~\ref{eq:idealLimit} denotes the full BSM effect which generally includes also the interference between SM and BSM amplitudes and is thus not necessarily positive.

\begin{figure}[tb]
\centering
\includegraphics[width=0.48\textwidth]{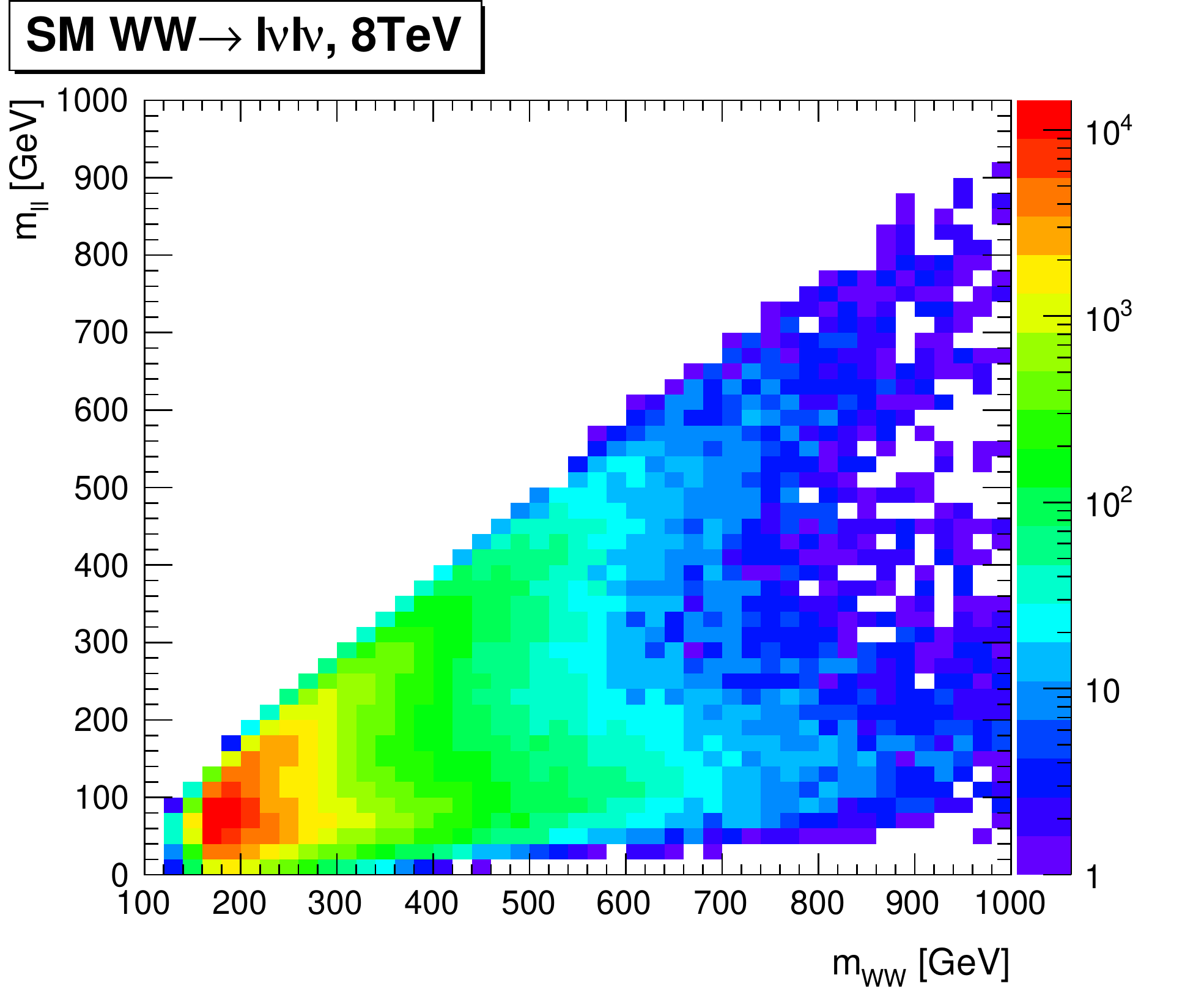}\quad
\includegraphics[width=0.48\textwidth]{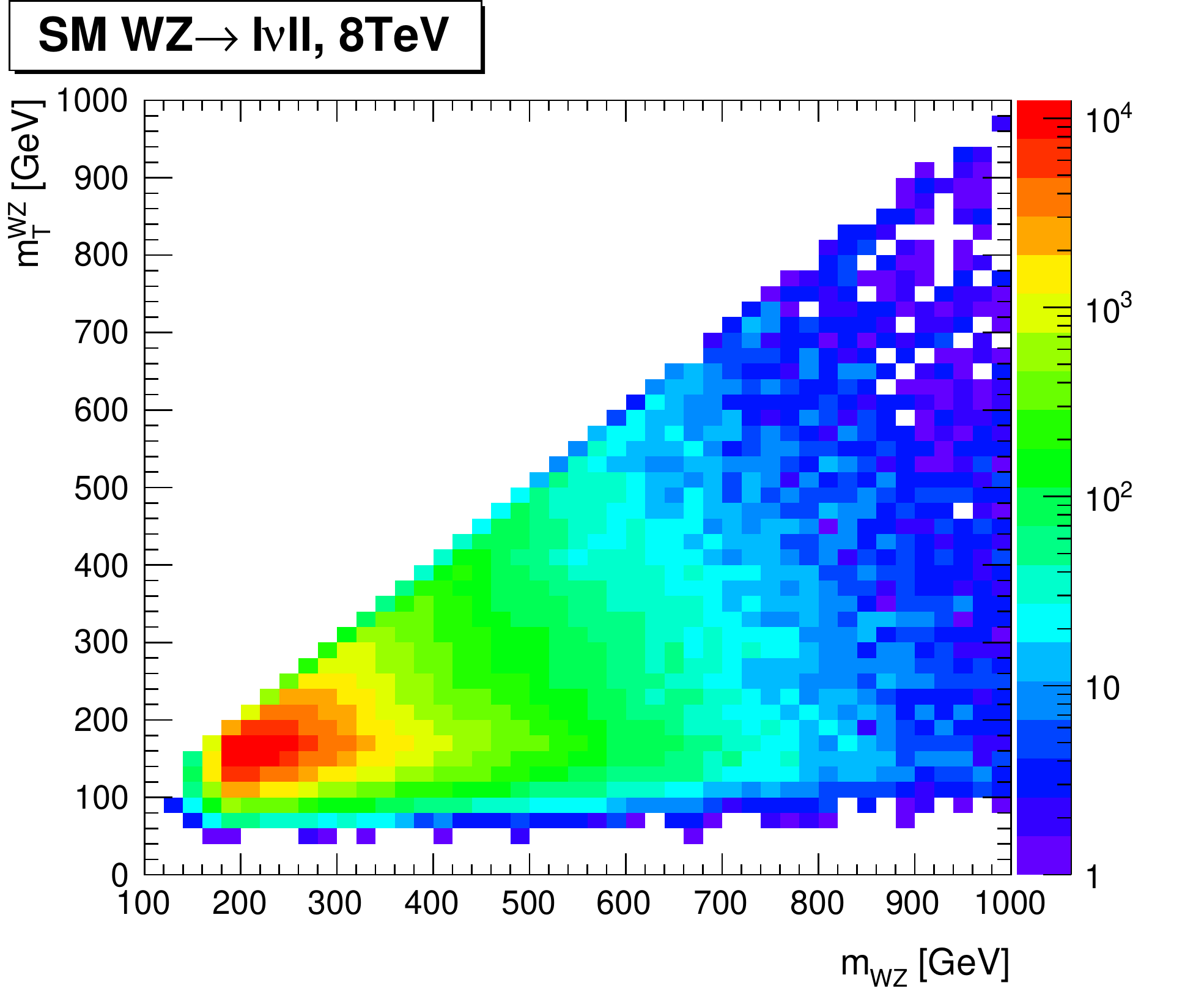}
\quad 
\caption{Left: Event distribution in the plane of the invariant mass of dilepton system, $m_{\ell\ell}$  (which is reported by the experiment) versus $m_{WW}$ (which corresponds to $\sqrt{\hat{s}}$.). Right: Similar plot for $m_T^{WZ}$ vs $m_{WZ}$. Both histograms are based on $5\times 10^5$ events.}
\label{fig:mwztSmwz}
\end{figure}

However, in realistic analyses this approach is limited by the incapability of reconstructing the full invariant mass of the diboson system when one or both gauge bosons decay into neutrinos.\footnote{The ATLAS analysis at $\sqrt{s} =$ 7 TeV does consider the full reconstruction of $m_{WZ}$~\cite{Aad:2012twa} but  the $m_{WZ}$ resolution is low due to the low resolution on $E_T^{\rm{miss}}$.}
In this case other observables, which we generically denote as $M_{vis}$, are constructed from the available information in the final state. For example, these can be the dilepton invariant mass $m_{\ell\ell}$ in the case of $WW$~\cite{Khachatryan:2015sga},  the transverse mass $m_{T}^{WZ}$ in the case of $WZ$ production~\cite{Aad:2016ett,Aaboud:2016yus}, or the transverse momentum of a gauge boson $p_T(V)$~\cite{Aad:2012twa}. 
The problem with this approach is that all these observables exhibit a poor correlation with the physically relevant scale $m_{VV}$, as can be seen from Fig.~\ref{fig:mwztSmwz} for $m_{\ell\ell}$ (left) and $m_{T}^{WZ}$ (right). A similar situation is present also for $p_T(V)$. As a consequence, the cut on $m_{VV}$ does not simply map onto a corresponding cut on $M_{vis}$:
\begin{equation}
 \int_0^{m_{VV}^{\rm max}} dm_{VV} \frac{d\sigma}{dm_{VV}} 
 \not\approx
  \int_0^{M_{vis}^{cut}} dM_{vis}\frac{d\sigma}{dM_{vis}}~,
\end{equation}
for any values of $M^{cut}_{vis}$.

\begin{table}[t]
\begin{center}\vspace{0.5cm}
\begin{tabular}{c|c|c|c|c|c|c} 
\hline \hline
 & \multicolumn{6}{c}{$m_{\ell\ell}$~(GeV)} \\
  & $<200$ & $<400$ & $<600$ & $<800$  & $<1000$ & $<\infty$ \\ 
\hline
$m_{WW}>400$~GeV & 0.54 & 0.85  & 0.97 & 0.99  & 1.0 & 1.0\\
$m_{WW}>600$~GeV & 0.43 & 0.65  & 0.87 & 0.97  & 0.99 & 1.0\\
$m_{WW}>800$~GeV & 0.36 & 0.59 & 0.71 & 0.90   & 0.97 & 1.0\\
$m_{WW}>1000$~GeV & 0.32 & 0.53 & 0.64 & 0.78   & 0.92 & 1.0  \\ \hline\hline

\end{tabular}
\caption{\label{tab:CutWW} Ratio of the number of events with and without $m_{\ell \ell}$ cut in the high $m_{WW}$ region, $N_{\rm evs}^{m_{\ell\ell}<m_{\ell\ell}^*}(m_{WW}>m_{WW}^*)/N_{\rm evs}(m_{WW}>m_{WW}^*)$, for the SM $p p \to W^+ W^- \to \ell \nu \ell \nu$ process at 8 TeV.}
\end{center}
\end{table}

Such a poor correlation implies that imposing a cut on $M_{vis}$ does not remove all -- or at least a significant fraction of -- the events from the region with $m_{VV} > m_{VV}^{\rm max}$, 
resulting in an inconsistent EFT interpretation. \admir{This can be directly observed in Table~\ref{tab:CutWW} which shows the ratio of the number of events with and without the upper cut on $m_{\ell\ell}$ in the high $m_{WW}$ region for  $p p \to W^+ W^- \to \ell \nu \ell \nu$ in the SM at 8~TeV.} \david{For example, a cut of $m_{\ell\ell} < 600$~GeV will still allow 87\% (64\%) of the original events with invariant masses $m_{WW} > 600 ~(1000)$~GeV. Given Fig.~\ref{fig:mwztSmwz}, we expect the situation to be even worse in the case of $m_T^{WZ}$.}
A very similar problem is present in the case of LHC dark matter searches within the EFT approach. Also in that case the invariant mass of the system is not observable due to the missing energy, and the available observables are, in general, poorly correlated with it \cite{Racco:2015dxa}.

In this situation one can still set conservative bounds on the EFT parameters, imposing the EFT cut $m_{VV}^{\rm max}$ only on the simulated BSM events (not on the SM) and comparing with the observed events. A simple way to understand this approach is to simplify the $\chi^2$ analysis by approximating that the $68\%$CL bound comes from comparing the measured cross section in a given bin of the experimentally accessible distribution, $\sigma_{obs} \pm \Delta \sigma$, with the expected one, $\sigma_{\SM} + \sigma_{\rm BSM}$, and requiring the latter to be within the experimental error, namely
\be
	\sigma_{obs} - \Delta \sigma < \sigma_{\SM} + \sigma_{\rm BSM} < \sigma_{obs} + \Delta \sigma~.
	\label{eq:simplifiedchi2}
\ee
By applying the $m_{VV}$ cut on the BSM events, at the simulation level, we split $\sigma_{\rm BSM} =\sigma_{\rm BSM}^{m_{VV}<m_{VV}^{\rm max}} + \sigma_{\rm BSM}^{m_{VV}>m_{VV}^{\rm max}}$. 
If both these terms are positive (or both negative\footnote{The inequality of Eq.~(\ref{eq:ineq}) in this case is switched and a similar discussion applies. The procedure presented in Section~\ref{sec:recast} to set conservative bounds works for either sign (both positive or both negative) when no significant excess is observed.}) and as long as no significant excess is observed, then from the inequalities in eq.~\eqref{eq:simplifiedchi2} follows
\be
	\sigma_{obs}-\sigma_{\SM} - \Delta \sigma <  \sigma_{\rm BSM}^{m_{VV}<m_{VV}^{\rm max}}  < \sigma_{obs}-\sigma_{\SM} + \Delta \sigma~.
	\label{eq:ineq}
\ee
Under the above-mentioned assumptions, the resulting constraint on $\sigma_{\rm BSM}^{m_{VV}<m_{VV}^{\rm max}}$ provides a conservative bound on the EFT parameters, with the first inequality trivially satisfied.

Note that the positivity  assumption is not necessarily realized in general. The BSM contributions are schematically given by
\begin{equation}
   \sigma_{\rm BSM} \propto \left ( \mathcal{A}_{\SM}^* \mathcal{A}_{\rm BSM} + h.c. \right ) + |\mathcal{A}_{BSM}|^2 ~, 
\end{equation}
and can be negative if the interference terms dominates and is negative.
However, as discussed in the previous section, in the parameter space where the BSM effects are large enough to be observable, the quadratic terms typically dominate the low-energy part of the cross section where the EFT approach is reliable. Assuming also dim-8 contribution to be sub-leading implies $\sigma_{\rm BSM}^{m_{VV}<m_{VV}^{\rm max}}$ is positive. Furthermore, for large invariant masses (where the EFT is no longer valid) one would naively expect that the interference effect in this region is generally small due to a relatively small $\mathcal{A}_{\SM}$, which may justify assuming a positive $\sigma_{\rm BSM}^{m_{VV}>m_{VV}^{\rm max}}$. This can be explicitly seen in the right panel of Fig.~\ref{fig:xsec_aTGC}, which shows how the quadratic terms dominate in the high invariant mass region.

In case an excess is observed, hinting a possible new resonance, the above strategy fails to provide a reasonable bound. For instance, while the EFT cross section with the cut, $\sigma_{\rm BSM}^{m_{VV} < m_{VV}^{\rm max}}$, excludes the events beyond $m_{VV}^{\rm max}$, the data, $\sigma_{\rm obs}$, would include the entire contribution including those from the resonance region, leading to an unphysical fit of the EFT coefficients. This issue can be fixed \martin{choosing a larger confidence level interval so that the the lower limit in Eq. 2.5 is zero.} \david{The downside of this is that any information about the excess would get lost.}
Nonetheless, in our analysis we will not worry about this point anymore, since no significant excess has been observed in the available data.\footnote{\martin{Small fluctuations in a few bins are not expected to invalidate the analysis. If the tension becomes significant, these fluctuation will generate TGC bounds incompatible with zero at $68\%$CL (generated by a positive LHS in Eq.~\eqref{eq:ineq}). In that case one should simply choose a larger CL (e.g. $90\%$CL) where the tension disappears, so that the TGC bounds obtained from Eq.~\eqref{eq:ineq} are reliable.}}

\subsection{On the interference between SM and BSM amplitudes}
\label{subsec:noninterference}
We observed in our numerical study in Section~\ref{subsec:xsecWWandWZ} the suppression of the interference between SM and dim-6 operators (relative to the dim-6 squared contributions). Recently, Ref.~\cite{Azatov:2016sqh} showed that a rich theoretical structure behind this numerical observation can be revealed in the explicit computation of the helicity amplitudes. We summarize in this section the main results we have obtained applying such an analysis to diboson production processes.

Table~\ref{table:HelicityAmplitude} shows our results for the helicity amplitudes (see Appendix~\ref{app:HelAmp} for details). Naively, one could expect all the SM helicity amplitudes to asymptote to a constant at large energies, and in the presence of aTGCs to grow as $E^2/m_W^2$. This expectation is however modified in most cases by additional $m_W/E$ factors suppressing either the SM or the BSM part. Indeed, there is only one helicity choice of the $W$'s where the interference is enhanced by $E^2/m_W^2$ (the one with two longitudinal $W$'s), whereas there are various helicity choices in which the quadratic terms are enhanced by $E^2/m_W^2$ or $E^4/m_W^4$.
This seems to be one of the reasons for a diminished sensitivity of LHC observables to the linear term in anomalous couplings observed in numerical simulations.
As long as the experimental precision does not allow one to probe small $\delta$, high energy bins will be sensitive mostly to the aTGC quadratic terms.
\begin{table}[t!]
\begin{center}
\scalebox{0.995}{
\begin{tabular}{ c|c|c|c }
\hline 
Helicity & $ \mathcal{A}_{SM}$ & $\mathcal{A}_{BSM}$  & $\sigma/(g^4_{SM}/E^2)$ \\[0.1cm]  \hline
$\mathcal{A}_{\psi\psi\rightarrow 00}$ & $\sim\mathcal{O}(g^2_{SM})$ & $\sim\mathcal{O}(g^2_{SM} E^2/m_W^2\, \delta )$ & $\sim 1+ (E^2/m^2_W) \delta + (E^4/m_W^4) \delta^2$  \\ [0.35cm]
$\mathcal{A}_{\psi\psi\rightarrow \mp\pm}$ & $\sim\mathcal{O}(g_{SM}^2)$ & $\sim\mathcal{O}(g^2_{SM}\, \delta)$ & $\sim 1+ \delta + \delta^2$  \\ [0.35cm] 
$\mathcal{A}_{\psi\psi\rightarrow \pm0}$ & $\sim\mathcal{O} \left( g^2_{SM} m_W/E \right )$ &  $\sim\mathcal{O}(g^2_{SM} E/m_W\, \delta)$ & $\sim m^2_W/E^2+ \delta + (E^2/m_W^2)\delta^2$  \\ [0.35cm]
$\mathcal{A}_{\psi\psi\rightarrow \pm\pm}$ & $\sim\mathcal{O} \left( g^2_{SM} m^2_W/E^2 \right )$ & $\sim\mathcal{O}(g^2_{SM} E^2/m_W^2\, \delta)$ & $\sim m^4_W/E^4 + \delta + (E^4/m_W^4)\delta^2$  \\ [0.3cm] \hline
\end{tabular}
} 
\caption{\label{table:HelicityAmplitude} 
Individual helicity contributions to diboson production cross section $\psi \psi \to W^+W^-$. $\delta$ is a short-hand notation for an appropriate linear combination of the anomalous couplings.
}
\end{center}
\end{table}

It turns out that the underlying principle behind the structure in Table~\ref{table:HelicityAmplitude} is due to
the helicity selection rule which forbids the interference between amplitudes with  different total helicities~\cite{Azatov:2016sqh}.
The detailed discussion and derivations are postponed to Appendix~\ref{app:NonInterferenceSMandBSM}. Here, we briefly summarise the most relevant results. We use the same notation and operator basis as in~\cite{Cheung:2015aba,Azatov:2016sqh} (the Warsaw basis~\cite{Grzadkowski:2010es}), in which operators with derivatives are removed in favour of those with more fields using the gauge bosons equations of motion.

We consider the following \david{classes of} $D=6$ operators that are relevant to  diboson production (similarly for the anti-holomorphic operators),
\begin{equation}\label{eq:dim6noFer}
c_1\, F^3~, \quad c_2\, \phi^2\, F^2~,\quad c_3\, (\phi D \phi)^2~, \quad c_4\, \bar{\psi}\gamma\psi\, \phi D \phi~,
\end{equation}
which include \david{${ Q}_{W}$, ${ Q}_{\phi WB}$, ${Q}_{\phi D}$, and ${Q}_{\phi\psi}$, respectively (in the notation of Ref.~\cite{Grzadkowski:2010es})}.
The normalization of $c_i$ is given in Eq.~(\ref{eq:eff_lagr_general}).

In the first (second) column of Table~\ref{table:NDA} we show the estimated sizes of the individual operator contributions to diboson production cross section $\psi \psi \to V V$ at linear (quadratic) order. The third column gives the typical collision energy ($E \sim \sqrt{\hat s}$) for which dim-6$\times$dim-6 dominates over SM$\times$dim-6 for a given operator. Clearly, the observed suppression in our fits of the interference term for the $F^3$ operator (corresponding to the aTGC $\lambda_z$) can be understood from the helicity selection rules.  One important consequence is that the energy scale above which the quadratic term dominates over the interference is suppressed by the factor $\sqrt{m_W/\Lambda}$. Thus, the energy range where the EFT is valid {\em and} the quadratic term dominates is larger than in a generic situation, and may be non-trivial even when the UV completion is weakly coupled.   

On the other hand, both linear and quadratic terms are suppressed in $\phi^2 F^2$ \martin{and $(\phi D \phi)^2$} whereas no suppression is present in $\bar{\psi}\gamma\psi\, \phi D \phi$.
Estimating $c_{2,4} \sim g_*^2$, the energy range where the quadratic terms dominates is non-trivial only for strongly coupled UV completions where $g_* \gg g_{\rm SM}$.
The estimated sizes of the individual operator contributions in Table~\ref{table:NDA}  match the explicit computations summarized in  Appendix~\ref{app:NonInterferenceSMandBSM}.

\begin{table}[t!]
\begin{center}
\scalebox{1.0}{
\begin{tabular}{ c|c|c|c }
\hline 
$\mathcal{O}_i$ & $\sigma_{SM\times dim_6}/(g^4_{SM}/E^2)$ & $\sigma_{dim_6^2}/(g^4_{SM}/E^2)$  & Energy range for $\sigma_{dim_6^2} > \sigma_{SM \times dim_6}$ \\[0.1cm]  \hline
$F^3$ & $\ds\frac{c_1}{g_{SM}}\frac{m^2_W}{\Lambda^2}$ & $\ds\frac{c_1^2}{g^2_{SM}}\frac{E^4}{\Lambda^4}$ & $ \Lambda \sqrt{\ds\frac{m_W}{\Lambda}} \left (\ds\frac{g_{SM}}{c_1} \right)^{1/4} < E < \Lambda$ \\ [0.37cm]
$\phi^2 F^2$ & $\ds\frac{c_2}{g^2_{SM}}\ds\frac{m_W^2}{\Lambda^2}$ & $\ds\frac{c_2^2}{g^4_{SM}}\frac{m_W^2 E^2}{\Lambda^4}$ & $\Lambda \left ( \ds\frac{g_{SM}}{\sqrt{c_2}} \right ) < E < \Lambda$ \\[0.37cm]
$(\phi D \phi)^2$ & $\ds\frac{c_3}{g^2_{SM}}\frac{m_W^2}{\Lambda^2}$ & $\ds\frac{c_3^2}{g_{SM}^4}\frac{m_W^4}{\Lambda^4}$ & $-$ \\[0.37cm]
$\bar{\psi}\gamma\psi\, \phi D \phi$ & $\ds\frac{c_4}{g^2_{SM}}\ds\frac{E^2}{\Lambda^2}$  & $\ds\frac{c_4^2}{g^4_{SM}}\ds\frac{E^4}{\Lambda^4}$ & $\Lambda \left ( \ds\frac{g_{SM}}{\sqrt{c_4}} \right ) < E < \Lambda$ \\[0.37cm] \hline
\end{tabular}
} 
\caption{\label{table:NDA} 
Individual operator contributions to diboson production cross section $\psi \psi \to V V$ at linear (first column) and quadratic (second column) order. Third column shows the energy range for which dim-6$\times$dim-6 dominates over SM$\times$dim-6 for a given operator.
}
\end{center}
\end{table}

An important remaining issue is the size of the interference between SM and dim-8 operators, which is formally of the same order in the EFT expansion as dim-6$\times$dim-6. In what follows, we restrict ourselves to the models in which dim-6$\times$dim-6 contribution dominates over SM$\times$dim-8. We leave a detailed study of SM$\times$dim-8 for future work.

\section{Facilitating the EFT interpretation of existing searches}
\label{sec:recast}

In order to show the impact of the EFT validity cuts discussed above on the aTGC extraction from real data, we recast two $8~ \TeV$ analysis: CMS $W^+W^-$~\cite{Khachatryan:2015sga} with $ \cL = 19.4~ \fb^{-1}$ and ATLAS $W^\pm Z$~\cite{Aad:2016ett} using $\cL = 20.3~ \fb^{-1}$; as well as the recent $13~\TeV$ analysis of $W^\pm Z$ production by ATLAS~\cite{Aaboud:2016yus}, using $\cL = 3.2~\fb^{-1}$.

In all cases leptonic decays of the $W$ and $Z$ are considered, leading to dilepton and trilepton signals. Since these are the most sensitive channels, neglecting the other (hadronic) ones should not qualitatively impact the combined results.
The extraction of aTGC bounds in the EFT approach will be carried out with the prescription described in Section~\ref{sec:EFTLimitSetting}.

In addition to the analyses mentioned above, ATLAS~\cite{Aad:2016wpd} (CMS~\cite{CMS:2013qea}) also measured  $W^+W^-$ ($W^\pm Z$) process in the fully leptonic channel at $\sqrt{s}=8$ TeV, using the full data set. The analysis by ATLAS uses the transverse momentum of the leading lepton ($p_T^{\rm{lead}}$) to set limits on aTGC whereas the CMS result has not been interpreted as the limit on aTGC. We opt not to recast these searches here. Once again, adding them to our analysis should not change our result significantly, and it is not crucial for our purpose of discussing how to set bounds on aTGC consistently within the EFT approach.
For similar reasons, we also do not recast the analyses using the data at $\sqrt{s} = 7$ TeV.

We implement aTGC using \textsc{FeynRules}~\cite{Alloul:2013bka}  in a \textsc{UFO} model \cite{Degrande:2011ua}, which is then imported in \textsc{MadGraph}5~\cite{Alwall:2014hca} to simulate our signal events. The signal events are further parton-showered and hadronizied by \textsc{Pythia8}~\cite{Sjostrand:2014zea}. 

\subsection{$W^+W^-\to \ell\nu_{\ell} \ell'\nu_{\ell'}$}
\label{subsec:WWlvlv}
The CMS analysis of the $W^+W^-\rightarrow l^+\nu l^-\bar{\nu}$ process at $\sqrt{s} = 8$ TeV provides the differential cross section in terms of the invariant mass of the dilepton system ($m_{\ell\ell}$)~\cite{Khachatryan:2015sga}. The analysis includes four event categories, defined in terms of the number of jets and lepton flavor.

\david{Following the experimental selection, we keep} only events with two oppositely charged isolated leptons with different flavor. The selected leptons are required to have $p_T(l) > 20$ GeV and $|\eta(l)| < 2.5 (2.4)$ for electrons (muons). A lepton is declared to be isolated if the $p_T$-sum of all particles within the isolation cone size $R_{iso} = 0.3$, excluding the lepton itself, is less than 10\% of the $p_T(l)$. The dilepton system is further restricted to satisfy $p_T(\ell \ell) > 30$ GeV and $m_{\ell \ell} > 12$ GeV. The remaining particles in an event are clustered into anti-$k_T$ jets with $R_{jet} = 0.5$ using the \textsc{FastJet} package~\cite{Cacciari:2011ma}. The reconstructed jets are required to have $p_T(j) > 30$ GeV and $|\eta(j)|<4.5$. The events with more than one reconstructed jet are vetoed. The missing transverse momentum, $\vec{E}^{\rm miss}_T$, is defined as the negative vector sum of $p_T$ of all reconstructed particles in the event. The projected $E^{\rm miss}_T$ is defined as the component of $\vec{E}^{\rm miss}_T$ transverse to the nearest lepton if $\Delta \phi (l,\, \vec{E}^{\rm miss}_T) < \pi/2$, otherwise the projected $E^{\rm miss}_T$ is simply defined as $|\vec{E}^{\rm miss}_T|$. We demand that projected $E^{\rm miss}_T$ is bigger than 20 GeV.

Our procedure successfully reproduces the number of events of $q\bar{q}\rightarrow W^+W^-$ for different lepton flavors in both zero and one-jet category (see Table 4 of Ref.~\cite{Khachatryan:2015sga}) within few \% discrepancy, validating our analysis.\footnote{The two same-flavor categories are rather difficult to validate as the analysis uses DY MVA as one of the cuts.} Since the $gg\rightarrow W^+W^-$ process \david{represents only a $\sim 3\%$ contribution to the total cross section, the aTGC contribution arising from it} is certainly sub-leading. Therefore, for simpicity, in our analysis we simulate only $q\bar{q}\rightarrow W^+W^-$ and simply rescale it to match  the total contribution from both processes.


\begin{figure}[tb]
\centering
\includegraphics[width=0.315\textwidth]{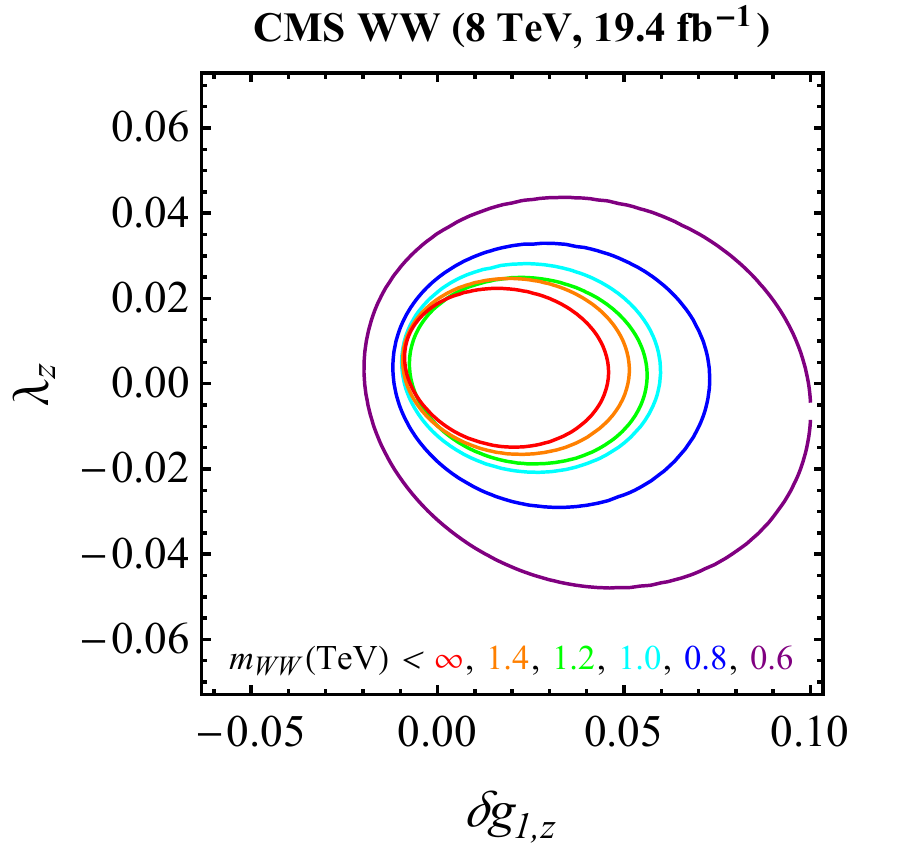}\quad
\includegraphics[width=0.31\textwidth]{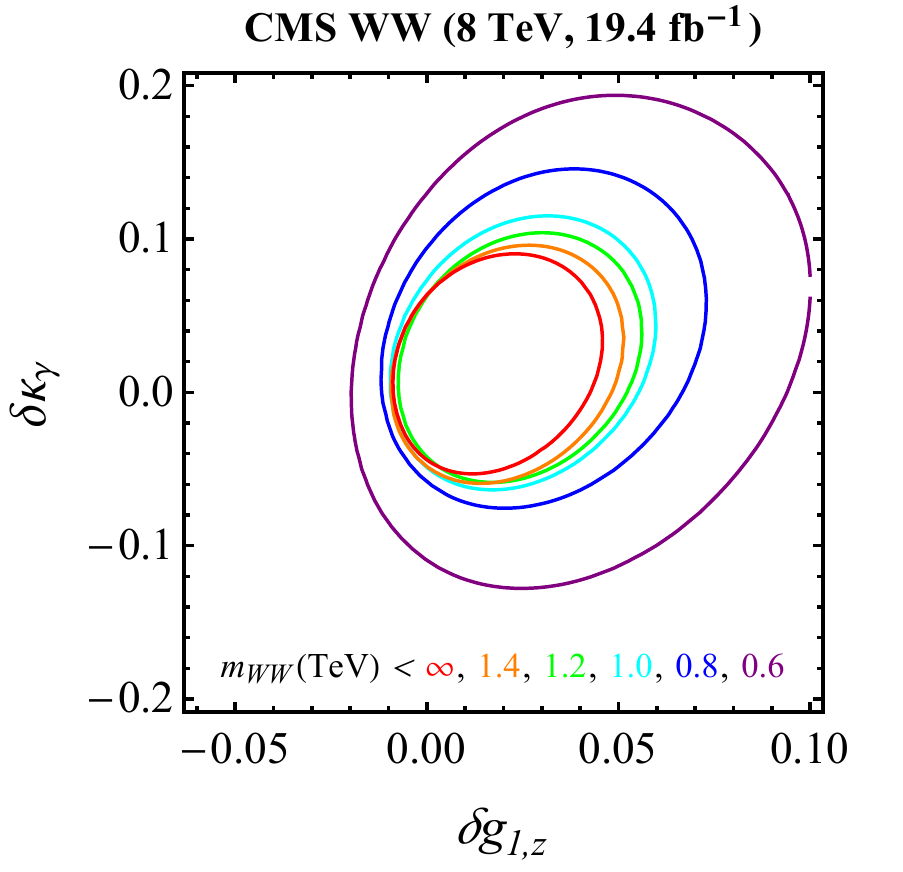}\quad
\includegraphics[width=0.314\textwidth]{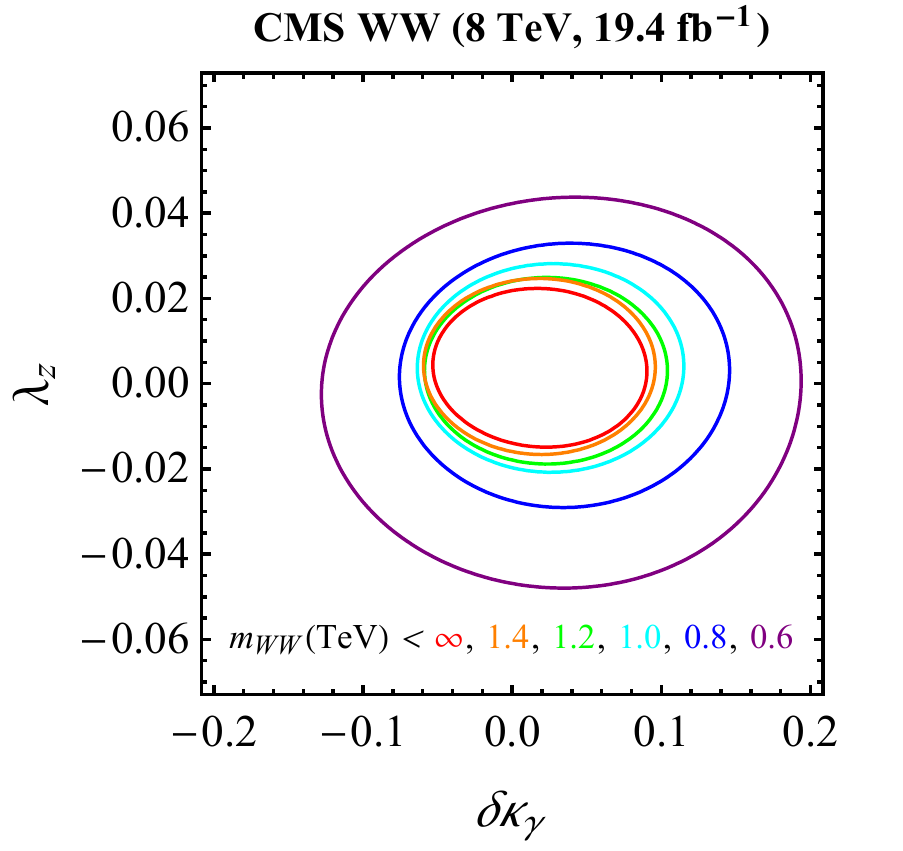}
\caption{$68\%$ CL region from 8~TeV CMS $pp \to W^+W^-$ searches for different $m_{WW}$ cuts.}
\label{fig:wwlimits}
\end{figure}


In order to recast the analysis as a limit on aTGC, we extract from Fig.~4 of Ref.~\cite{Khachatryan:2015sga} the observed number of events, the expected SM contribution to $W^+W^-$, and the total SM background in the $m_{\ell\ell}$ distribution.
As was discussed in Section~\ref{sec:EFTLimitSetting}, the upper cut $m_{WW}^{\rm max}$ is imposed only on the BSM part at the simulation level, $\Delta\sigma^{(i)}_{{\rm BSM, } m_{WW}^{\rm max}} \equiv \Delta\sigma^{(i)}_{\rm BSM} \left (\sqrt{\hat{s}} < m_{WW}^{\rm max} \right )$, to get a conservative bound.
The resulting cross section $\Delta \sigma^{(i)}$, where $i$ runs over eight $m_{\ell\ell}$ bins, is given by
\begin{equation}\label{eq:TotalSigmaWW}
  \Delta \sigma^{(i)} = \Delta \sigma^{(i)}_{\rm SM\, MC} \left( 1+ \frac{\Delta\sigma^{(i)}_{{\rm BSM,} m_{WW}^{\rm max}} }{ \Delta \sigma^{(i)}_{\SM}} \right) ~,
\end{equation}
where the cross section is rescaled such that our SM prediction matches the Monte Carlo results of Fig.~4 of Ref.~\cite{Khachatryan:2015sga}.
Note that $\Delta \sigma^{(i)}_{\rm BSM}$ in Eq.~\ref{eq:TotalSigmaWW} includes the interference between SM and BSM amplitudes (linear terms in aTGC), as well as quadratic terms in aTGC:
\be
	 \frac{\Delta\sigma^{(i)}_{{\rm BSM, } m_{WW}^{\rm max}} }{ \Delta \sigma^{(i)}_{\SM}}  =  B^{(i)}_a \kappa_a +  C^{(i)}_{ab}\kappa_a \kappa_b~,
\label{eq:ww-bin-bsm}
\ee
where $a$ and $b$ run over the three aTGC $\kappa \equiv \left\{ \lz, \dgz, \dkg\right\}$. In order to solve for $B^{(i)}$ and $C^{(i)}$, one needs to run the simulation for at least ten points in the aTGC parameter space. We perform a profile likelihood fit to the binned $m_{\ell\ell}$ distribution.\footnote{The numerical approach adopted here is explained in more details in Ref.~\cite{Greljo:2015sla} in the context of electroweak Higgs production analyses.} 

The resulting sensitivity on the aTGC are illustrated in Fig.~\ref{fig:wwlimits}, where in each plot we show the 68\% CL limit on two aTGC profiling the likelihood over the third one, for values of $m_{WW}^{\rm max}<$~$\infty$ (red), 1.4 (orange), 1.2 (green), 1.0 (cyan), 0.8 (blue) and 0.6~TeV (purple).
As was expected, the sensitivities are weakened as the cut is lowered. 
However, the dependence of the limits on the EFT cut is small up to $m_{WW}^{\rm max} \simeq 1~ \TeV$ and becomes important only for lower cutoffs. This implies that the bounds on aTGC obtained from the $8~\TeV$ WW searches without any cutoff offer approximately valid constraints for new physics scenarios with mass scales above $\sim 1~ \TeV$, as long as dim-8 contributions can be neglected. 
Interestingly enough, even for a relatively small $m_{WW}^{\rm max}$, the obtained limits are rather competitive with respect to those from the combined fit to Higgs and LEP2 data~\cite{Falkowski:2015jaa}.
Finally, it is worth mentioning that the aTGC bounds that we obtain without any $m_{WW}$ cut are in a good agreement with the limits quoted by the experimental collaboration~\cite{Khachatryan:2015sga} and by Ref.~\cite{Butter:2016cvz}.


\begin{figure}[tb]
\centering
\includegraphics[width=0.45\textwidth]{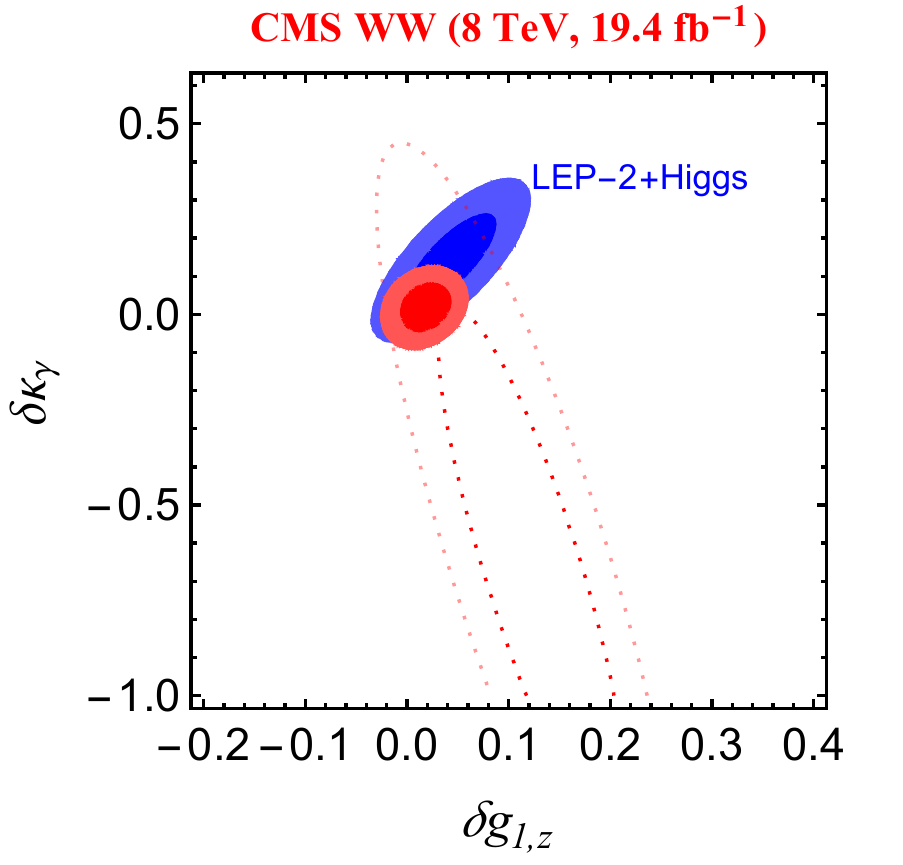}
\caption{Recast of the CMS analysis of $W^+W^-\rightarrow l\nu l\nu$ process at $\sqrt{s} = 8$ TeV and $19.4~\fb^{-1}$~\cite{Khachatryan:2015sga}. Bounds on the anomalous triple gauge couplings obtained expressing the signal strengths in each bin up to quadratic (red-filled) and linear (red-dashed) order in aTGC, respectively. No cuts on truth  $m_{WW}$ are applied.}
\label{fig:wwlimitsQuadvsLinear}
\end{figure}

In Fig.~\ref{fig:wwlimitsQuadvsLinear} we compare the sensitivities obtained from recasting the CMS $8~ \TeV$ WW analysis by including (red-filled) or excluding (red-dashed) quadratic terms in dim-6 operators. We observe that the limits are much weakened when only linear terms are included, in agreement with the discussion of Section~\ref{subsec:xsecWWandWZ}. Therefore, in BSM scenarios where quadratic dim-6 and linear dim-8 terms are of the same size (following the general EFT counting), the latter are expected to generate similar changes in the aTGC bounds. This implies that the aTGC bounds derived by including quadratic dim-6 terms largely overestimate the constraints for such BSM scenarios.
Let us also note that non-included QCD NLO corrections might change qualitatively the interference terms, since the LO terms happen to be suppressed. \martin{This is in fact confirmed by preliminary results shown in Ref.}~\cite{Azatov:talk}. Therefore, the result of the linear fit in Fig.~\ref{fig:wwlimitsQuadvsLinear} should be taken with caution, but the main message (large sensitivity to quadratic corrections) is not affected by this caveat.

This is unlike the limits from Higgs+LEP2 combined dataset~\cite{Falkowski:2015jaa} where the linearized fit (shown in blue) leads to similar results as the one including quadratic corrections. In fact, the observables of this analysis (Higgs signal strengths and $e^+e^-\to W^+W^-$ differential cross section) receive large SM contribution and the dominant new physics effect occurs at order $\Lambda^{-2}$ due to the interference of dim-6 operators with the SM.

Nonetheless, as mentioned before, in a large class of models,  dim-6 squared terms dominate over linear dim-8 in the low-energy EFT. In these situations  stronger bounds from the quadratic fit can be applied. This class of models includes, but is not \martin{necessarily} limited to\footnote{\martin{In principle one should be able to engineer a (non strongly coupled) model where the various free parameters are fine-tuned so that the relevant dim-8 Wilson Coefficients are suppressed. 
}}, strongly coupled models.

\subsection{$W^\pm Z\to  \ell^\pm \nu_{\ell}\ell^+ \ell^-$}
\label{sec:WZrecast}

In the ATLAS $W^\pm Z$ analyses (both 8 and 13 TeV)~\cite{Aad:2016ett,Aaboud:2016yus}, limits on aTGC are derived from the transverse mass spectrum of the $W Z$ system ($m_T^{WZ}$) imposing no upper cut on the momentum transfers in the process. 

We use \textsc{MadGraph}5 to generate parton-level events for a set of points in aTGC parameter space. 
The fiducial phase space region is defined with the following set of cuts (see also Table~1 of Ref.~\cite{Aad:2016ett}). Three isolated charged leptons with $\eta(\ell) < 2.5$ are required, two of which must form a pair with opposite charge and same flavor to reconstruct the $Z$ boson (with $|m_{\ell\ell}-m_Z| <  10$ GeV), while the third is associated with the $W$ decay.  While the leptons from the $Z$ decay need to pass the cut $p_T(\ell^\pm_Z) > 15$ GeV, the lepton from the $W$ decay is required to have $p_T(\ell_W)>20$ GeV. The separation between leptons is required to be $\Delta R (\ell^+_Z \ell^-_Z)>0.2$ and $\Delta R (\ell^\pm_Z \ell_W)>0.3$, respectively. Finally, the $W$ transverse mass needs to satisfy $m_T^W >30$ GeV. The same set of cuts has also been applied in the 13 TeV analysis.

We perform an analysis equivalent to the $WW$ case to set aTGC limits for different $m_{WZ}$ cuts. In the 8 TeV search, we focus on the measured $m_T^{WZ}$ differential cross section in the fiducial phase space, reported in Fig.~5 of Ref.~\cite{Aad:2016ett}.\footnote{~Equivalently, one could also recast the reconstructed event distribution in $m_T^{WZ}$ in Fig.~12 of~\cite{Aad:2016ett}. We, however, encourage experimental collaboration to continue publishing unfolded differential distribution measurements which can be  more accurately included in our analyses (using theorist level tools).} We first reproduce the SM predictions for the cross sections in each bin after applying the overall NLO QCD $K$-factor from Ref.~\cite{Grazzini:2016swo}, which serves as a check of our simulation procedure. We then compute for each bin the linear and quadratic dependence of the cross section on the aTGC as in Eq.~\eqref{eq:ww-bin-bsm} with same upper cuts on $m_{WZ}$ on the BSM events as discussed for $WW$. We use $\Delta \sigma^{{\rm fid}\, (i)} / \Delta \sigma^{{\rm fid}\, (i)}_{\rm{SM}}$ measurements to construct the $\chi^2$ as function of the three aTGC. Theoretical errors due to the limited SM predictions shown with an orange band in Fig.~5 of Ref.~\cite{Aad:2016ett} are added in quadrature to the experimental errors. In the case of the 13 TeV analysis, we use the number of observed and expected events in $m_{T}^{WZ}$ bins shown in Fig.~1 of Ref.~\cite{Aaboud:2016yus}. 

\begin{figure}[tb]
\centering
\includegraphics[width=0.33\textwidth]{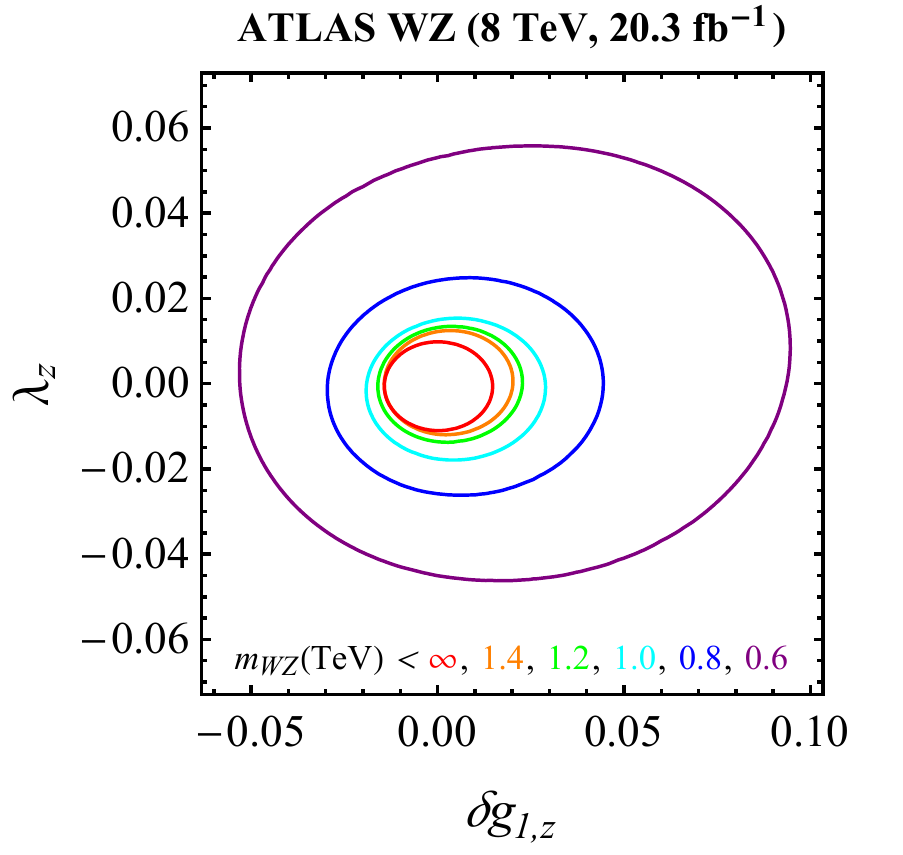}
\includegraphics[width=0.32\textwidth]{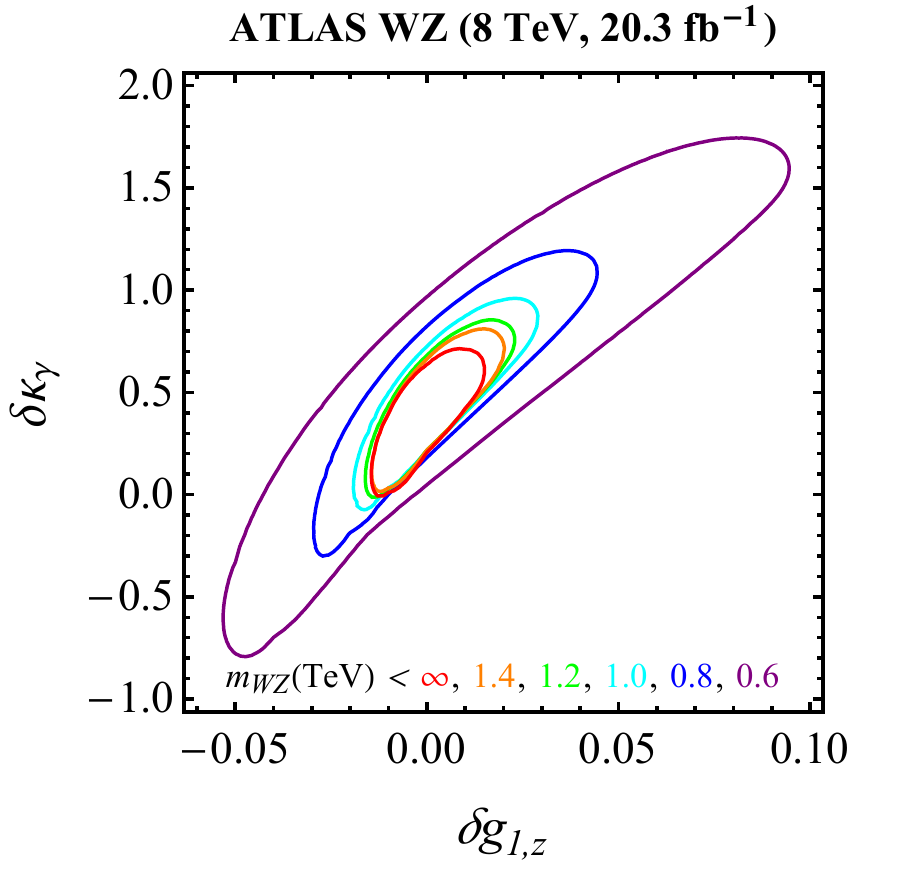}
\includegraphics[width=0.325\textwidth]{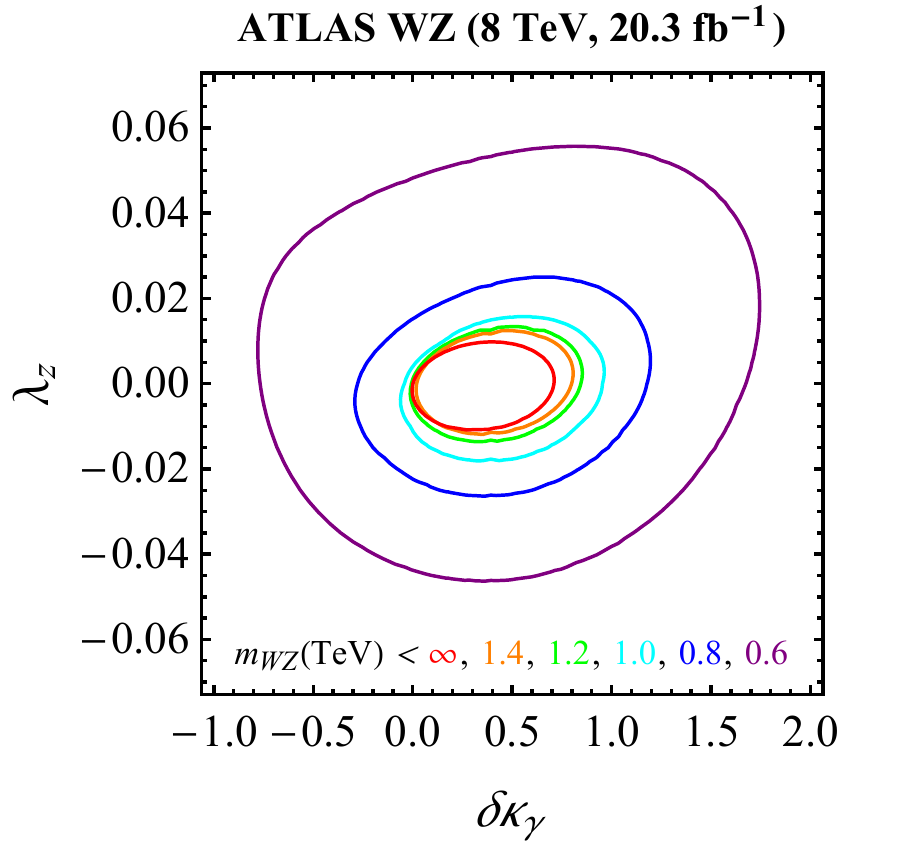}
\caption{$68\%$ CL region from 8 TeV ATLAS $pp \to W^\pm Z$ searches for different $m_{WZ}$ cuts.}
\label{fig:wzlimits}
\end{figure}

\begin{figure}[tb]
\centering
\includegraphics[width=0.33\textwidth]{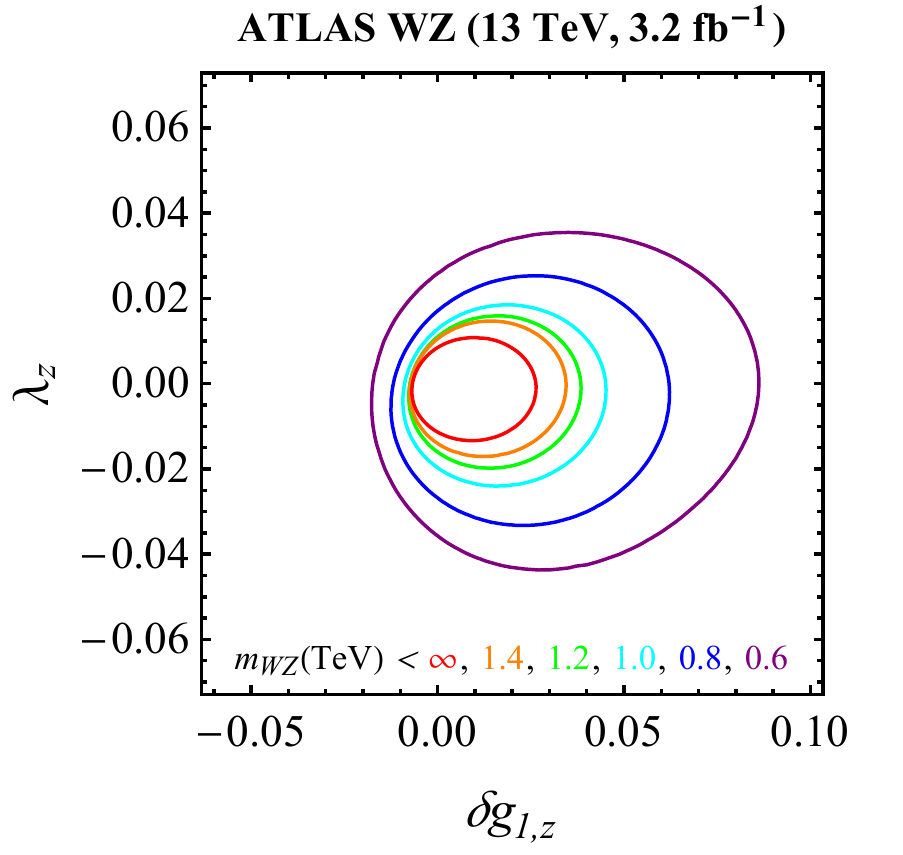}
\includegraphics[width=0.32\textwidth]{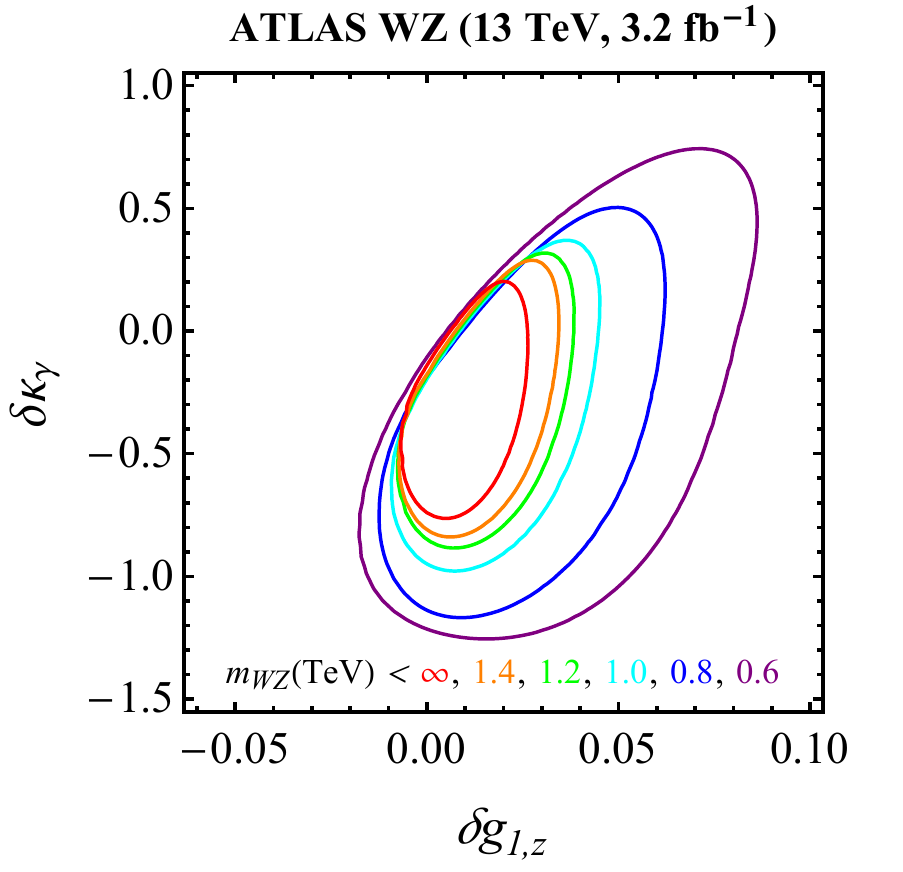}
\includegraphics[width=0.325\textwidth]{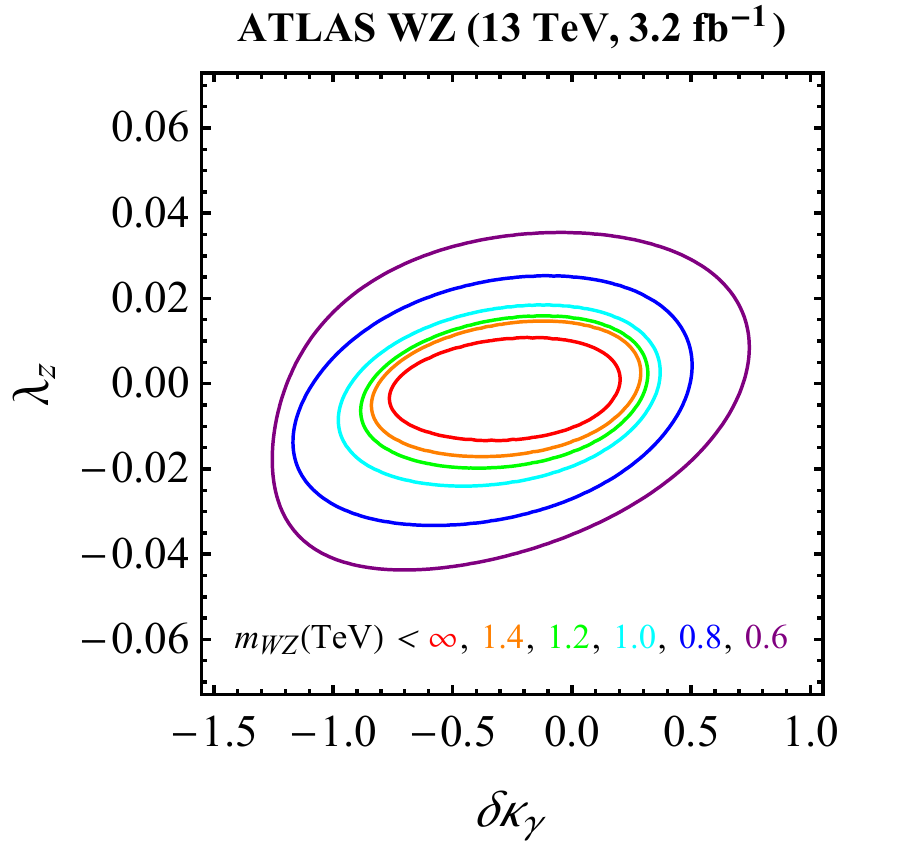}
\caption{$68\%$ CL region from 13 TeV ATLAS $pp \to W^\pm Z$ searches  for different $m_{WZ}$ cuts.}
\label{fig:wz13_limits}
\end{figure}

The 68\%~CL region resulting from the fit of the 8 TeV (13 TeV) data are presented in Fig.~\ref{fig:wzlimits} (Fig.~\ref{fig:wz13_limits}) in two-dimensional aTGC planes after profiling over the third parameter. Same $m_{VV}$ cuts are imposed as in the $WW$ case, with the same color-code. 
 The situation is very similar to that discussed for WW. Our limits from the 8 TeV analysis without any $m_{WZ}$ cut are in a good agreement with the limits quoted by the experimental collaboration and by Ref.~\cite{Butter:2016cvz}. As expected, the limits on aTGC soften with a tighter cut, albeit only to a small degree up to $m_{WZ}^{\rm max} \sim 1~\TeV$. We have checked  that also in this channel the (strong) aTGC limits are mainly due to large quadratic terms ($C^{\rm (bin)}_{ab}$) in Eq.~\eqref{eq:ww-bin-bsm}, and thus assume implicitly negligible contributions from linear dim-8 terms.

\begin{figure}[tb]
\centering
\includegraphics[width=0.33\textwidth]{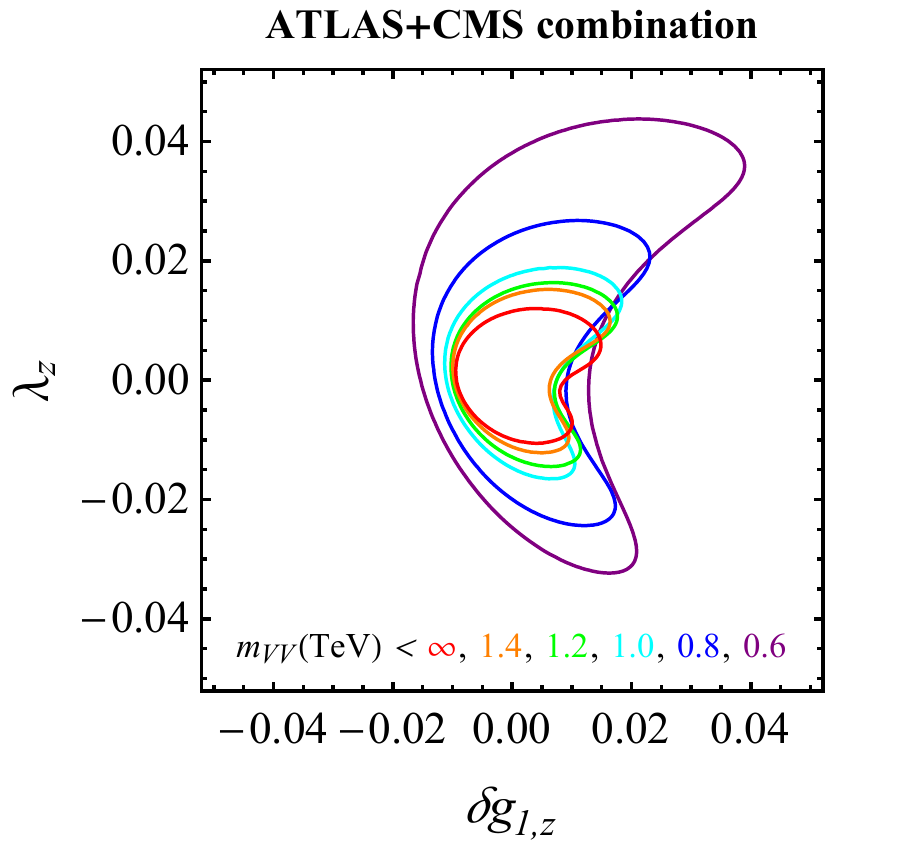}
\includegraphics[width=0.32\textwidth]{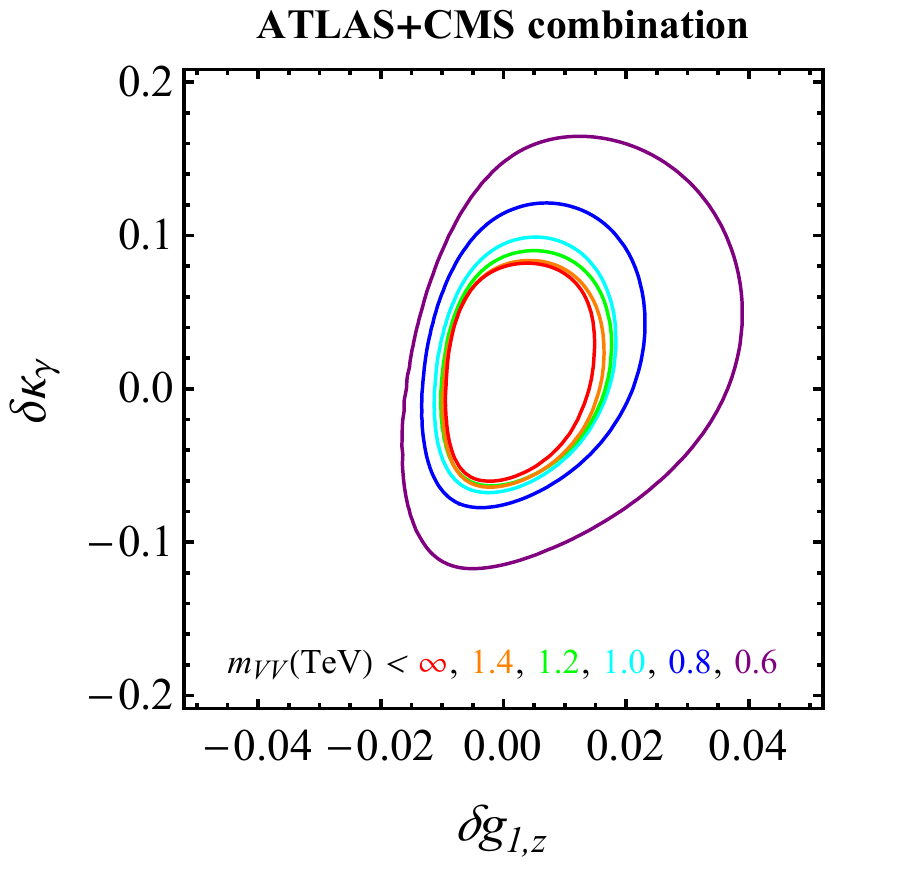}
\includegraphics[width=0.325\textwidth]{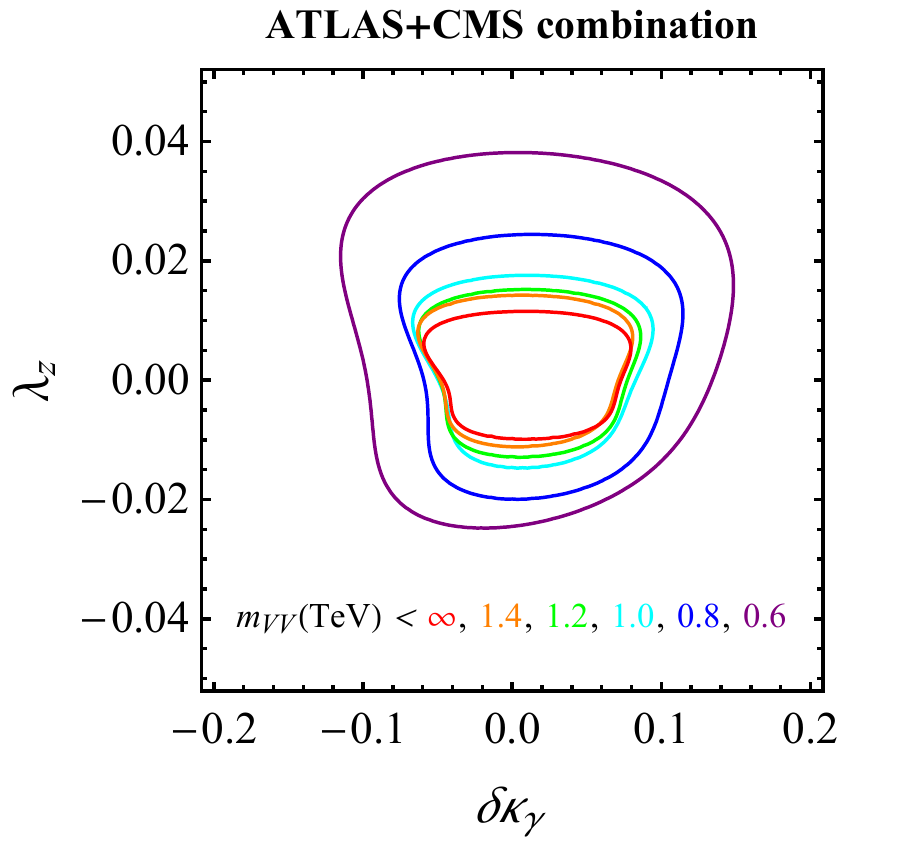}
\caption{Combined $68\%$ CL region from CMS $WW$ (8~TeV) and ATLAS $WZ$ (8+13~TeV) searches  for different $m_{VV}$ cuts.}
\label{fig:comb_limits}
\end{figure}

\subsection{Combination}
\label{sec:combination}

\begin{table}[t]
\begin{center}\vspace{0.5cm}
\begin{tabular}{c|c|c|c|c|c|c} \hline

$m_{VV}^{\rm max}$ & $\infty$ & 1400 & 1200 & 1000 & 800 & 600\\ 
 (GeV) &  &  &  &  &  & \\ \hline
$\delta g_{1,z} (\%)$ & $[-1.2,~ 2.0]$ & $[-1.2,~ 2.2]$ & $[-1.3,~ 2.4]$ & $[-1.4,~ 2.5]$ & $[-1.7,~ 3.2]$ & $[-2.1,~ 5.4]$\\
$\delta \kappa_\gamma ~ (\%)$ & $[-7.8,~ 9.9]$ & $[-8.3,~ 10]$ & $[-8.4,~ 11]$  & $[-9.0,~ 11]$ & $[-10,~ 15]$ & $[-15,~ 21]$\\
$\lambda_z ~ (\%)$ & $[-1.3,~ 1.3]$ & $[-1.5,~ 1.7]$ & $[-1.8,~ 1.8]$  & $[-2.1,~ 2.1]$ & $[-2.9,~ 3.0]$ & $[-4.2,~ 4.8]$ \\ \hline\hline

\end{tabular}
\caption{\label{tab:95bound} Profiled $95\%$ CL bounds on the each aTGC from CMS $WW$ (8~TeV) and ATLAS $WZ$ (8+13~TeV) searches for different $m_{VV}$ cuts.}
\end{center}
\end{table}

In Fig.~\ref{fig:comb_limits} we combine the limits from the three analysis described above, CMS $WW$ at 8 TeV~\cite{Khachatryan:2015sga}, ATLAS $WZ$ at 8 TeV~\cite{Aad:2016ett} and ATLAS $WZ$ at 13~TeV~\cite{Aaboud:2016yus}, showing the combined 68\% CL region in the three aTGC as a function of the EFT cut on $m_{VV}$, where $V=W,Z$. The 95\% CL bounds on each single aTGC after profiling over the other two, for different $m_{VV}$ cuts, are shown in Table~\ref{tab:95bound}.

Since the present sensitivity on the aTGC is driven by the quadratic terms, the final likelihood is not expected to be a Gaussian. For this reason we encourage the experimental collaborations to present separately the $68\%$ and $95\%$ CL contours in the three 2-dimensional aTGC planes shown above.

\section{An explicit model testing the EFT approach }
\label{sec:model}

\adam{
The goal of this section is to evaluate the validity of the EFT description of VV production for a specific example of a UV model that replaces the EFT for $E \geq \Lambda$.}
Given a concrete model with new  particles, we can constrain it via two different procedures.
One is to directly calculate the model's predictions for VV production and to confront them with the experimental data so as to constrain the parameter space (the masses and couplings) of the BSM model. 
Alternatively, one could first integrate out the new particles and calculate the Wilson coefficients of the low-energy EFT as a function of the masses and couplings of the model.   
Then constraints on the model's parameter space can be obtained by recasting the constraints on the EFT parameters derived in Section~\ref{sec:combination}. 
We expect that the two procedures should give the same results when the new particles are heavy enough (and if dim-8 terms are negligible), and different results  when they are so light as to be produced on-shell at the LHC. 
The energy scale below which the two procedures diverge sets the validity range of the SMEFT for that particular new physics scenario.

We are interested in a model  where the aTGC $\delta g_{1,z}$ is generated at tree-level in the low-energy EFT without large contributions to other electroweak precision observables. 
The latter requirement is non-trivial.
Indeed, $\delta g_{1,z}$ can be generated by integrating out new heavy vector bosons mixing with W and Z bosons. 
However, the mixing generically also shifts the Z and W boson couplings to fermions, as well as the W boson mass, which were accurately measured in LEP and other precision experiments. 
In the model below, the absence of large corrections to the electroweak precision observables will be achieved by a fine-tuned cancellation between contributions from different heavy vectors.  

We consider the SM extended by the following degrees of freedom:
\bi 
\item a vector triplet $V_\mu^i$, $i=1\dots 3$ transforming as an adjoint under the SM $SU(2)_L$;
\item a vector field $V_\mu^0$ which is a singlet under the SM gauge group. 
\ei 
For simplicity, we are assuming the triplet and the singlet have the same mass $m_V$. The interactions between the new vectors and the SM are given by 
\bea
\begin{split}
{\cal L}  & \supset   {i \over 2}  g_L \kappa_H V_\mu^0  H^\dagger  \overleftrightarrow{D_\mu} H 
+ g_L V_\mu^0  \sum_{f \in \ell,q}  \kappa_{f} Y_f  \bar f \bar \sigma_\mu  f 
+ g_L V_\mu^0  \sum_{f \in e,u,d}  \kappa_{f} Y_{\bar f^c} f^c \sigma_\mu  \bar f^c
\\
&  + {i \over 2}  g_L \kappa_H'  V_\mu^i H^\dagger \sigma^i  \overleftrightarrow{D_\mu} H 
+ {g_L \over 2} V_\mu^i \sum_{f \in \ell, q}  \kappa_{f}'   \bar f \sigma^i \bar \sigma_\mu  f,
\end{split}
\eea  
where  $H^\dagger   \overleftrightarrow{D_\mu} H =  H^\dagger {D_\mu} H - D_\mu H^\dagger  H$. 
\adam{ This is not a UV complete model, as it introduces new vector fields without an associated gauge symmetry. However, it can be easily embedded in a UV complete framework. 
For example, the masses could arise in a perturbative framework of deconstruction \cite{ArkaniHamed:2001ca} where the SM electroweak symmetry is replicated, and the larger group is broken to the SM via a VEV of a bi-fundamental (``link" ) scalar fields. 
Alternatively, the vectors could be composite excitations of a strongly interacting sector with a global $SU(2)\times U(1)$ symmetry weakly gauged by the SM electroweak bosons,
as in composite Higgs models \cite{Kaplan:1983fs}.
The following discussion does not depend on how the model is UV completed.} 

To derive the low-energy EFT of this model at tree-level, one integrates out the heavy vectors by solving their equation of motion and plugging the solution back to the Lagrangian. With this procedure one obtains the following $D$=6 operators in the  effective Lagrangian: 
\bea 
\label{eq:frankenstein_leff}
\begin{split}
\cL_{\rm eff}  &= \cL_{\rm SM} 
-  {g_L^2 \over 8 m_V^2} \Big ( 
 i  \kappa_H'  H^\dagger \sigma^i  \overleftrightarrow{D_\mu} H 
 +  \sum_{f \in \ell, q}  \kappa_f'  \bar f \sigma^i \bar \sigma_\mu  f \Big )^2 
 \\ & 
 -  {g_L^2 \over 8 m_V^2} \Big ( 
 i  \kappa_H  H^\dagger \overleftrightarrow{D_\mu} H 
-   \sum_{f \in \ell, q}  \kappa_{f}  Y_f  \bar f \sigma^i \bar \sigma_\mu  f  
 -   \sum_{f \in e,u,d} \kappa_{f} Y_{\bar f^c} f^c \sigma_\mu  \bar f^c  
  \Big )^2 
 + \cO(m_V^{-4})~.
\end{split}
\eea 
With a bit of algebra, one can massage these operators to a form that fits one of the $D$=6 bases in the literature. For example,  in the Warsaw basis  the Wilson coefficients are found to be:\footnote{
We use the original operator normalization of Ref.~\cite{Grzadkowski:2010es}, we however absorb the EFT scale $\Lambda$ into the Wilson coefficients, $c_i/\Lambda^2 \to \bar c_i/v^2$ ($v\approx 246$ GeV).} 
 \bea
 \label{eq:frankenstein_WB}
 \bar c_{H \Box}  & = & - \left ( {3 \over 2} \kappa_H^{\prime 2} + {1 \over 2} \kappa_H^{2} \right )  {m_W^2   \over  m_V^2}, \qquad 
  \bar c_{H D}  =  - 2 \kappa_H^{2}  {m_W^2   \over  m_V^2}, \qquad  
\bar c_{H} = - 4 \lambda \kappa_H^{\prime 2}   { m_W^2  \over m_V^2},
 \\
\, [\bar c_{Hf}]_{IJ} & = &  -\sqrt 2 \kappa_H^{\prime 2} {m_f \over v}  {m_W^2 \over m_V^2} \delta_{IJ},  \quad 
\,  [\bar c_{Hf}^{(3)}]_{IJ}  = - \kappa_H'  \kappa_f'  { m_W^2  \over  m_V^2}  \delta_{IJ}, \quad 
\,  [\bar c_{Hf}^{(1)}]_{IJ}  = 2 \kappa_H \kappa_{f}  Y_f   {m_W^2  \over m_V^2}  \delta_{IJ}, 
\nonumber
\eea
plus a set of four-fermion operators.

The parameters  $\kappa_H$, $\kappa_H'$, $\kappa_f$, and $\kappa_f'$  characterize the coupling strength of the new vectors to the SM and  are a-priori free parameters. 
In the following, the couplings to fermions are assumed to be flavor universal and diagonal. Moreover, we assume that they are related to the couplings to the Higgs field as  
\beq
\label{eq:frankenstein_tuning}
\kappa_f' = -{g_L^2 \over 2 g_Y^2 }{\kappa_H^2 \over  \kappa_H'}, 
\qquad 
\kappa_f  =  - {\kappa_H \over 2}. 
\eeq 
One can show that this tuning ensures that the couplings of the light gauge boson eigenstates (identified with the SM gauge bosons) to the fermions are not shifted at tree level from their SM value.\footnote{There remains a correction to $G_F$ which, indirectly, also affects the measured value of the  gauge couplings to fermions. To get rid of it, one needs to invoke another fine-tuned UV contribution to the 4-fermion operator $(\bar \ell_1\bar \sigma_\mu \ell_2)(\bar \ell_2\bar \sigma_\mu \ell_1)$ responsible for the muon beta decay from which $G_F$ is extracted. For this reason we do not consider its contribution to $\delta g_{1,z}$, even though according to the matching of Eq.~\eqref{eq:warsaw_atgc} it should be there.}

With these conditions imposed, the parameters space is 3-dimensional and can be  characterized by the couplings $\kappa_H$, $\kappa_H'$ and the mass $m_V$. 
The latter is approximately the mass of the two neutral and one charged heavy vector eigenstates, up to small corrections of order $v^4/m_V^4$. 
In the low-energy EFT below the scale $m_V$ one finds aTGCs of the SM gauge bosons described by\footnote{%
In this EFT there are also corrections to the Higgs couplings (which depend also on the combination $\kappa_H'/m_V$), but they are not important for the following discussion.} 
\beq
\label{eq:dg1zeft}
\delta g_{1,z} = - \kappa_H^2 {m_W^2 \over 2 s_\theta^2 m_V^2} + \cO(m_V^{-4}),   
\eeq  
while $\delta \kappa_\gamma = \lambda_z = 0$ at tree level. 
 Note that $\delta g_{1,z}$ is sensitive to the UV physics only via the combination  $\kappa_H/m_V$.  
Thus, for large $m_V$, diboson production at the LHC  is sensitive only to this particular combination, 
\adam{while the dependence on  $\kappa_H'$ cancels out after imposing the tuning conditions in \eref{frankenstein_tuning}}.  
On the other hand, for $m_V$ in the kinematic range of the LHC all the 3 parameters can be probed via  diboson production.  

\begin{figure}[th]
\bc 
\includegraphics[width=0.5 \textwidth]{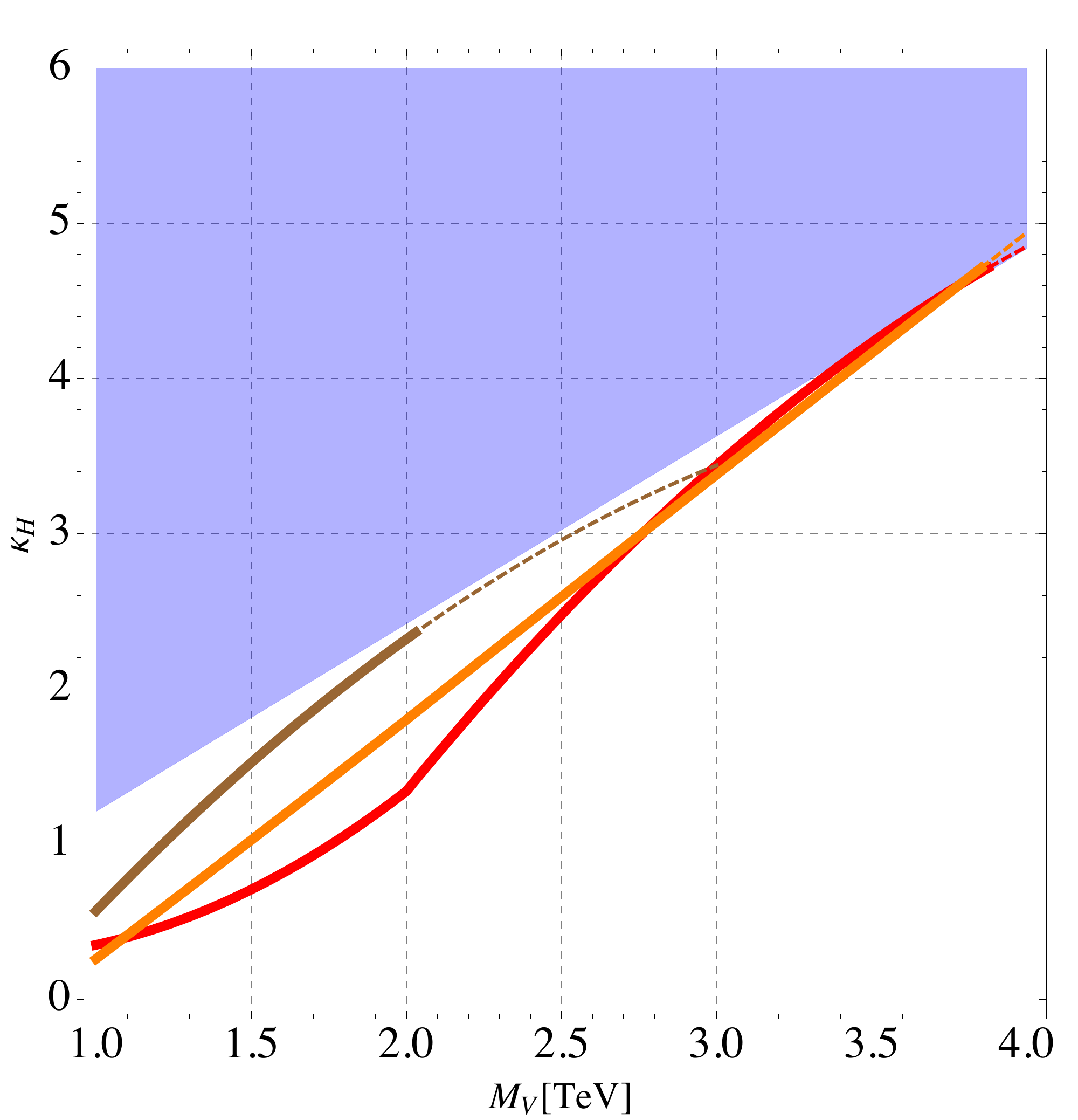}
\ec 
\caption{
Limits on the $\kappa_H$ coupling as a function of vector boson mass in the model discussed in this section. 
Different lines correspond to $\kappa_H' = 3 \kappa_H$ (red),  $\kappa_H' = \kappa_H$ (brown), and $\kappa_H' = -3 \kappa_H$ (orange).
The solid lines turn into  dashed ones at the scale when the UV model becomes non-perturbative, which we estimate as the scale where the total width of at least one of the heavy vectors exceeds $m_V/2$. 
The blue region is the parameter space excluded by recasting the EFT limits on $\delta g_{1,z}$ as limits on $\kappa_H/m_V$ using the matching in \eref{dg1zeft}. 
}
\label{fig:cms8ww_sawplot}
\end{figure}

We are ready to discuss the validity range of the EFT for the model described above. We will illustrate the quantitative determination of the validity range using as example the limits set by the CMS analysis of $W^+W^-$ production at $\sqrt{s} = 8$~TeV  \cite{Khachatryan:2015sga}. The results are summarized in \fref{cms8ww_sawplot}. We plot the direct limits on the parameter $\kappa_H$ as a function of $m_V$ for three different choices of the ratio $\kappa_H'/\kappa_H$. Since the aTGC in the leading-order SMEFT is independent of $\kappa_H'$ we expect that,  for large enough $m_V$, the limits are independent of that ratio. This is indeed the case for $m_V \gtrsim 3$~TeV. On the other hand, for  $m_V \lesssim 3$~TeV, when the new vectors enter the kinematic range of the  $\sqrt{s} = 8$~TeV LHC, the limits on $\kappa_H$ may easily vary by a factor of 2 depending on $\kappa_H'$.

In \fref{cms8ww_sawplot} we also show the parameter space excluded by recasting EFT limits on aTGCs  using \eref{dg1zeft} and the bounds obtained without any upper cut on $m_{WW}$. In this case, the limits, by construction, are independent of $\kappa_H'$. As expected, the EFT and the direct approach yield consistent limits  for $m_V \gtrsim 3$~TeV. Therefore, the scale of $3$~TeV is an approximate lower limit on the EFT cut-off $\Lambda$ such that, for this particular UV completion,  the SMEFT provides a valid description of diboson production at the $\sqrt{s} = 8$~TeV LHC. 

Note that, for this example, the true (direct) limits are always stronger than the ones derived indirectly by recasting the limits on the aTGC. 
Thus, the EFT approach provides a conservative limit on the parameters, even without restricting the kinematic range of experimental data used in the analysis.  

This example suggests that diboson measurements at the $\sqrt{s}=8$~TeV LHC can be adequately described using the SMEFT provided the EFT cut-off (or the scale of the BSM particles) is at least 3~TeV.  For $\sqrt{s}=13$~TeV LHC the necessary cut-off is expected to be even larger. For a lower cut-off, the limits on BSM models derived by recasting limits on the aTGC may have an order 1 error. Since the parameters of the low-energy EFT at the leading order depend on the cut-off as  $1/\Lambda^2$, they carry a large suppression factor for $\Lambda \gtrsim 3$~TeV. Given that the diboson measurements are currently sensitive to the aTGC of order 0.01, only rather strongly coupled UV theories can be efficiently constrained by the LHC using the EFT approach. This can be seen in \fref{cms8ww_sawplot}, where only $\kappa_H \gtrsim 3$ can be probed in the EFT validity regime of the LHC. Even larger couplings are needed if the aTGC are induced at the 1-loop level. Obviously, when the couplings are too large the UV model becomes non-perturbative, and then this particular description is no longer a useful UV completion. In this example the onset of a non-perturbative behavior occurs for $\kappa_H$ between 2 and 5, depending on $\kappa_H'$. Thus, the parameter window where the EFT description is useful is rather limited, at least for this particular UV completion.

\section{Conclusions}
\label{sec:conclusions}

On the one hand, it is well known that the EFT interpretation of (relatively) high-$p_T$ processes at the LHC -- such as diboson production, associated and VBF Higgs production, or even dark matter searches -- presents some challenges. On the other hand, the large amount of data gathered by the LHC on these processes, also considering the ever-increasing experimental and theoretical precision, has the potential to offer important insights on possible BSM scenarios, complementing the information obtained from LEP and precision low-energy experiments. In particular, this paper discusses in detail some of the most relevant challenges encountered while interpreting $WW$ and $WZ$ production at the LHC as measurements of anomalous triple gauge coupling in the context of the SMEFT. 

In principle, the leading BSM contribution to the relevant differential distributions should arise at $\mathcal{O}(\Lambda^{-2})$, due to the interference between SM and dim-6 operators. Next-to-leading corrections, of $\mathcal{O}(\Lambda^{-4})$, are instead due to dim-6 squared terms and interference of SM and dim-8 operators.
A consistent EFT analysis limited to dim-6 operators should therefore consider only interference terms as done in the Higgs+LEP-2 combined fit of Ref.~\cite{Falkowski:2015jaa}. In that case, including dim-6 squared terms does not qualitatively modify the results of Ref.~\cite{Falkowski:2015jaa}, which suggests a quick convergence of the EFT series and that $\mathcal{O}(\Lambda^{-4})$ terms can be neglected. However,  employing the same approach to diboson production at the LHC, we find very loose bounds in the linearized fit, and much stronger bounds when including quadratic terms (see Fig.~\ref{fig:wwlimitsQuadvsLinear}). For the latter, we agree with the conclusions of previous EFT fits to diboson production at the LHC (for example in~\cite{Butter:2016cvz}), as well as with the results quoted by experimental collaborations \cite{Aad:2016wpd,Khachatryan:2015sga}. In particular, including both interference and dim-6 squared terms we confirm that aTGC bounds from the LHC are stronger than the LEP ones. 

These results signify that the strong LHC bounds are dominantly due to the quadratic terms. 
In consequence, the  LHC limits on aTGCs  cannot be trivially combined with other datasets, since the likelihood is not approximately Gaussian. 
For this reason, we encourage experiments to publish the full likelihood function for the aTGCs (rather than just central values and errors),  which would allow theorists to easily perform a combination with other datasets and derive correct limits on BSM models.
The smallness of some of the interference terms at the LHC can be understood by an analysis of the relevant SM and BSM helicity amplitudes \cite{Azatov:2016sqh}, and is due to $\sim m_W / E$ suppression factors appearing in the SM or BSM part of the amplitude (see Section~\ref{subsec:noninterference}).
In this situation, LHC limits dominated by dim-6 squared terms can still be consistently interpreted in certain class of theories in which the dominant new physics contribution is indeed due to these terms, and in which dim-8 interference with the SM is also suppressed. For such theories, usually characterized by a several TeV mass gap from the SM and strong coupling, the LHC bounds as derived by experimental collaborations are applicable and indeed more stringent than the Higgs+LEP-2 ones, as recently pointed out in Ref.~\cite{Butter:2016cvz}.

Another handle on the validity of the EFT in LHC searches is to impose a cut on high-$p_T$ events, $m_{VV}^{\rm{max}} < \Lambda$, where $\Lambda$ is the assumed mass scale of the new physics, and perform the analysis for different values of $m_{VV}^{\rm{max}}$ (i.e. different assumptions on $\Lambda$). In this way, the EFT interpretation of the bounds for theories with a lower cut-off could also be possible. However, a complication arises due to the fact that the kinematical variable $m_{VV}$ can not be reconstructed experimentally if the final state includes neutrinos, in which case other kinematical variables such as $m_{\ell\ell}$ or $m_{T}^{WZ}$ are used to build differential cross sections. We find that these observables are very badly correlated with $m_{VV}$ (see Fig.~\ref{fig:mwztSmwz}), implying that a cut on them does not remove the unwanted high-$p_T$ events with a good enough efficiency. In this case, by imposing the desired $m_{VV}^{\rm{max}}$ cut at the simulation level on the BSM events only, consistent and conservative EFT bounds can still be obtained if no significant excess from the SM is observed. By recasting several CMS and ATLAS searches with this technique, for different values of $m_{VV}^{\rm{max}}$, we show that bounds with lower invariant mass cuts are in general less stringent (see Fig.~\ref{fig:comb_limits} and Table~\ref{tab:95bound}). In order to facilitate the interpretation of the measurement, we recommend presentation of the experimental results as a function of the EFT validity cut, $m_{VV}^{\rm{max}}$.

In order to explicitly check some of the conclusions from the EFT validity discussion, we introduce a simple BSM model generating aTGC at tree level, and compare the indirect bounds obtained from the EFT analysis (with no high-$p_T$ cut) with those obtained by directly analyzing the full model. We find that the EFT and direct bounds agree for resonance masses above $\sim 3~\TeV$. However, in this particular model, the EFT bounds are always more conservative than the direct ones, even down to masses of $\sim 1~\TeV$.

In conclusion, the bounds from the Higgs+LEP-2 global fit presented in Ref.~\cite{Falkowski:2015jaa} are applicable in the most general case in which SMEFT is well describing the underlying UV dynamics.
Instead, for a subset of theories (discussed in this work) that can also be matched to the SMEFT, $WW$ and $WZ$ searches at LHC provide the most stringent limits on aTGC.

\acknowledgments{
We thank Aleksandr Azatov,  Roberto Contino, and Francesco Riva for many insightful discussions.
A.F is partially supported by the ERC Advanced Grant Higgs@LHC and by the European Union�s Horizon 2020 research and innovation programme under the Marie Sklodowska-Curie grant agreements No 690575 and  No 674896.
 M.G.-A. is grateful to the LABEX Lyon Institute of Origins (ANR-10-LABX-0066) of the Universit\'e de Lyon for its financial support within the program ANR-11-IDEX-0007 of the French government. AG and DM are supported in part by the Swiss National Science Foundation (SNF) under contract 200021-159720. MS and AG thank the Mainz Institute for Theoretical Physics (MITP) while DM thanks the DESY Research Centre for hospitality during the completion of part of this work. 
}

\begin{appendix}
%

\section{Interference between SM and dim-6 BSM amplitudes}
\label{app:NonInterferenceSMandBSM}

In this appendix we present a discussion on the helicity structure of the SM and BSM amplitudes relevant to diboson production  {at the LHC}, the results of which are reported in Section~\ref{subsec:noninterference}. 

For completeness, we summarize some theoretical results about helicity amplitudes, following closely the discussion in Ref.~\cite{Azatov:2016sqh}. First, the little group scaling and Naive Dimensional Analysis uniquely relate the helicity of the three-point amplitude to the dimensionality of the coupling $g$ as $|h(A_3)| = 1 - [g]$ (see~\cite{Elvang:2013cua} for a review). In the SM, in the limit of unbroken electroweak symmetry, $h(A^{SM}_3)  = \pm 1$. Dimension-6 operators, such as the $F^3$ ones for example, can instead have $|h(A^{BSM}_3)| = 3$ since $[g] = -2$. It can also be shown \cite{Azatov:2016sqh} that all four-point SM amplitudes have vanishing total helicity, $h(A_4^{SM}) = 0$, in the massless limit except for an amplitude with four fermions involving both up- and down-quark Yukawa couplings.
In Refs.~\cite{Alonso:2014rga,Elias-Miro:2014eia} it has been shown that interesting results on the renormalization-group flow of dimension-6 operators can be obtained by considering their holomorphic properties. The same properties also help understanding the interference pattern with the SM.
In this context one defines the holomorphic and anti-holomorphic weights of an amplitude $A$ with $n(A)$ legs as  $w(A) = n(A) - h(A)$ and $\bar w(A) = n(A) + h(A)$. Then the total helicity of an $n$-point amplitude with an insertion of a higher-dimensional operator $\mathcal{O}$ is bounded as~\cite{Azatov:2016sqh}
\begin{equation}\label{eq:helicityBound}
  \bar{w}(\mathcal{O}) - n \leq h(A_n^{\mathcal{O}}) \leq n - w(\mathcal{O})~,
\end{equation}
where $w({\mathcal{O}}) = {\rm min}_A \{w(A) \}$, $\bar{w}({\mathcal{O}}) = {\rm min}_A \{ \bar{w}(A) \}$ are minimum weights over all the amplitudes induced by an operator $\mathcal{O}$. The helicity $h(A_n^{\mathcal{O}})$ in Eq.~(\ref{eq:helicityBound}) is even (odd) for even (odd) $n$. Lastly, two on-shell sub-amplitudes, $A_{m}, \, A_{m'}$, can combine to form an $n$-point amplitude, $A_n$, with $n = m + m' - 2$. The total helicity of the resulting amplitude is simply the sum of the helicities of the sub-amplitudes, namely, $h(A_n) = h(A_m) + h(A_{m'})$.

We now apply this formalism to understand the leading energy behavior of diboson ($WW$ or $WZ$) production cross sections in the presence of dimension-6 operators beyond the SM. We use the same notations and operator basis as in~\cite{Cheung:2015aba,Azatov:2016sqh}, namely, the Warsaw basis~\cite{Grzadkowski:2010es}. The relevant dimension-6 operators  are listed in Eq.~\eqref{eq:dim6noFer}.

\begin{figure}[t]
\centering
\minipage{0.48\textwidth}
  \includegraphics[width=.9\linewidth]{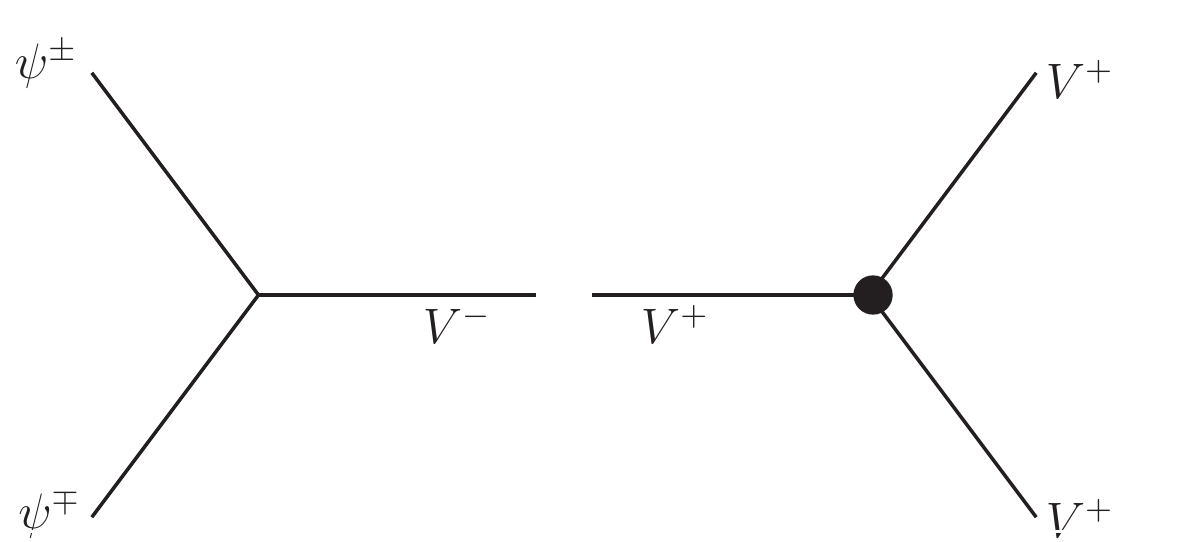}
\endminipage
\caption{The four-point amplitude involving $F^3$. The $F^3$ operator (shown as a dot) contributes to the three-point amplitude with the total helicity of 3 (right part of the diagram). The superscripts, $+$, $-$, denote the helicity of $\psi$ or $V$.}
\label{fig:four-pt-FFF}
\end{figure}

We start our survey with the $F^3$ operator, which  has the weight $(w,\, \bar{w}) = (0,\, 6)$ and contributes to the three-point amplitude $VVV$ with the helicity of 3. 
In the Warsaw basis there is one such operator, denoted as $O_{3W}$, contributing to the diboson production. It connects to SM three-point amplitude with a gauge boson and fermion pair, $\psi^\pm \psi^\mp V^-$, to form the four-point amplitude with the total helicity of 2, $h(A^{BSM}_4)=2$, as is illustrated in Fig.~\ref{fig:four-pt-FFF}. The SM four-point amplitude, instead, has a vanishing total helicity, $h(A_4^{SM})=0$. Therefore, in the massless limit (unbroken EW symmetry), the helicity selection rule forbids the interference between the SM and the BSM amplitude due to the $F^3$ operator. A non-vanishing interference thus requires two helicity flips. For instance, the SM amplitude $\psi^\pm \psi^\mp \phi\phi$ with $h=0$ can flip two helicities by connecting with two SM three-point vertices $V^+ \phi \phi$ (one $\phi$ gets a non-zero vacuum expectation value -VEV-) in order to interfere with the BSM amplitude $\psi^\pm \psi^\mp V^+ V^+$ with $h=2$. The two helicity flips imply a total suppression factor of $(m_W/E)^2$. The quadratic dim-6 term, on the other hand, does not require any helicity flip. The power counting of the cross section is given by 
\begin{equation}\label{eq:TTTT}
\begin{split}
 \sigma_{\psi\psi \rightarrow TT(++)} 
 &\sim \frac{g_{SM}^4}{E^2} \Big [ {m_W^4 \over E^4} + \underbrace{\frac{c_1}{g_{SM}} \frac{m_W^2}{\Lambda^2}}_{SM\times F^3} + \underbrace{\frac{c_1^2}{g^2_{SM}} \frac{E^4}{\Lambda^4}}_{F^3\times F^3} + \cdots \Big  ]~,
\end{split}
\end{equation}
where  {$c_1/\Lambda^2$ multiplies the $F^3$ operator in the Lagrangian}, the subscript $T$ refers to the transverse mode of gauge bosons, and the subscript $++$ specifies the helicities of the two gauge bosons (the same result holds for the $--$ helicities). 
 {In the Warsaw basis,  all other operators lead to a softer energy dependence of this helicity cross section. 
Clearly, the quadratic term grows rapidly with energy, while the interference term does not.}
For the energy range
\begin{equation}
 \Lambda \sqrt{\frac{m_W}{\Lambda}} \left (\frac{g_{SM}}{c_1} \right )^{1/4} < E < \Lambda~,
\end{equation}
the EFT is valid and the dim-6$\times$dim-6 contribution dominates over the interference term.
As discussed in \cite{Biekoetter:2014jwa,Contino:2016jqw}, $c_1 \gg g_{SM}$ (which may occur when the UV completion contains large couplings) increases the range where the quadratic term dominates over the interference one within the EFT validity regime.
For these particular diboson helicities, the relative suppression of the interference term has the effect of widening  that range by the factor of $(m_W/\Lambda)^{1/2}$. 
As a result, the quadratic term may dominate that helicity cross section even for $c_{1} \lesssim g_{SM}$, as long as $m_W \ll \Lambda$.

\begin{figure}[t]
\centering
\minipage{0.48\textwidth}
  \includegraphics[width=.9\linewidth]{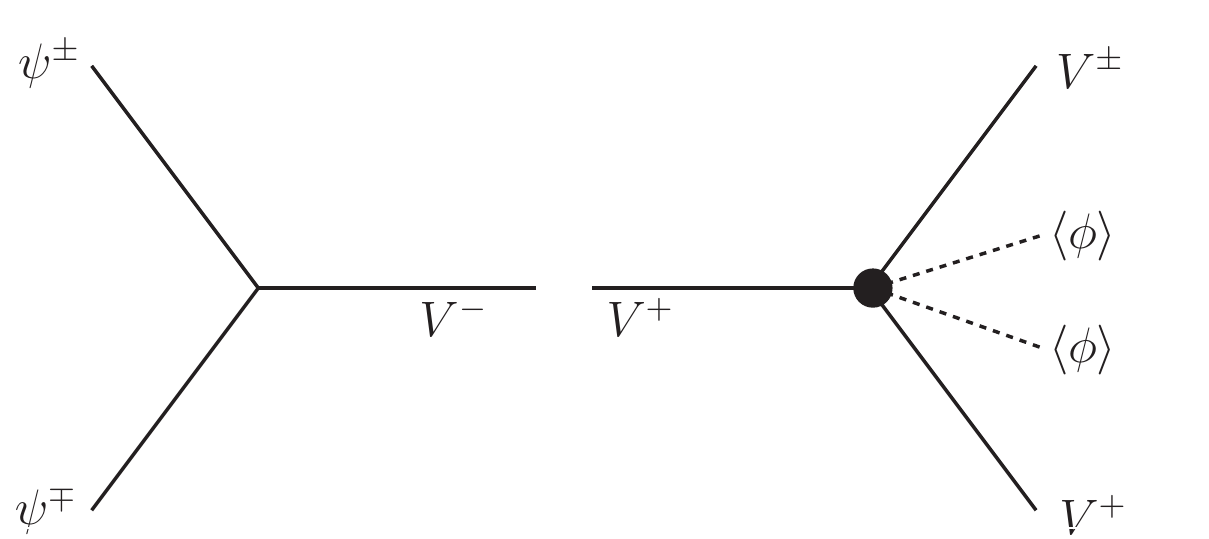}
\endminipage
\minipage{0.48\textwidth}
  \includegraphics[width=.9\linewidth]{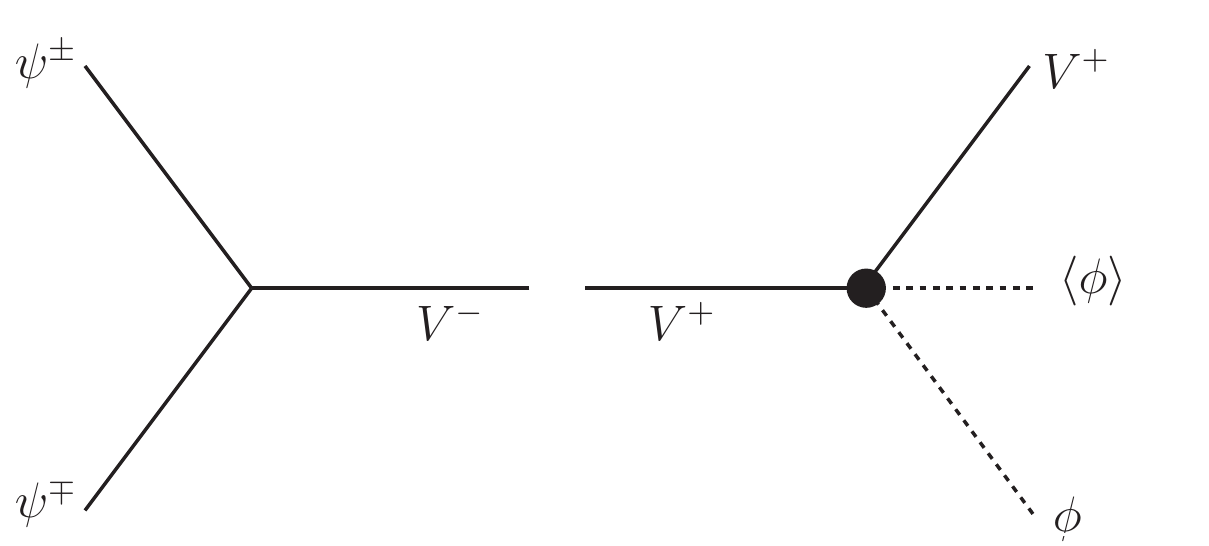}
\endminipage
\caption{The four-point amplitudes involving $\phi^2 F^2$. The $\phi^2 F^2$ operator (shown as dots) can contribute to the five-point amplitude with the helicity of 1 or 3 (right part of the left diagram). Similarly for the three-point amplitude with total helicity of 2 (right part of the right diagram).}
\label{fig:four-pt-ssFF}
\end{figure}

Let us now consider the $\phi^2 F^2$ operator with $(w,\, \bar{w}) = (2,\, 6)$. In the Warsaw basis, one such operator, denoted as $O_{HWB}$, contributes to diboson production. According to Eq.~(\ref{eq:helicityBound}),  $\phi^2 F^2$  can contribute to the four-point amplitude $V^+V^+\phi \phi$ with the helicity of 2 or to the five-point amplitudes $V^\pm V^+V^+ \phi\phi$ with the helicity of 1, 3. The five-point amplitudes can induce the three-point amplitudes with $h = 1,\, 3$ by replacing all two $\phi$'s with their VEVs, $\langle \phi \rangle = v$, as is seen in the right part of the left diagram in Fig.~\ref{fig:four-pt-ssFF}. The resulting four-point amplitudes, made by connecting them to SM three-point amplitude, have $h(A_4^{BSM}) = 0,\, 2$. The contribution of the case with $h(A_4^{BSM}) = 2$ to the cross section $\sigma_{\psi\psi\rightarrow TT(++)}$ is sub-leading, compared to Eq.~(\ref{eq:TTTT}), whereas the case with $h(A_4^{BSM}) = 0$ has a different  {energy} dependence, compared to Eq.~(\ref{eq:TTTT}) (note different helicities though, $+-$ vs. $++$),
\begin{equation}\label{eq:TTTT:hhFF}
 \sigma_{\psi\psi \rightarrow TT(+-)}
 \sim \frac{g_{SM}^4}{E^2} \Big [ 1 + \underbrace{\frac{c_2}{g^2_{SM}} \frac{m_W^2}{\Lambda^2}}_{SM\times \phi^2 F^2} + \underbrace{\frac{c_2^2}{g^4_{SM}} \frac{m_W^4}{\Lambda^4}}_{\phi^2 F^2 \times \phi^2 F^2} + \cdots  \Big  ]~,
\end{equation}
where we used $m_W \sim g_{SM} v$. One can also show that no other dimension-6 operator contributes terms growing with energy to this helicity cross section, therefore,  sensitivity of this final state to the EFT parameters is limited. On the other hand, the four-point amplitude $V^+V^+\phi \phi$ (right sub-diagram in Fig.~\ref{fig:four-pt-ssFF}) can contribute to the triple gauge vertex by replacing one of $\phi$ with its VEV. The resulting four-point amplitude, shown in Fig.~\ref{fig:four-pt-ssFF}, has the total helicity of 1 (therefore it requires one helicity flip). It contributes to the cross section $\sigma_{\psi\psi\rightarrow TL}$ whose parametric behavior is given by
\begin{equation}\label{eq:TTTL}
\begin{split}
 \sigma_{\psi\psi \rightarrow TL} 
 &\sim \frac{g_{SM}^4}{E^2} \Big [ {m_W^2 \over E^2} + \underbrace{\frac{c_2}{g^2_{SM}} \frac{m_W^2}{\Lambda^2}}_{SM\times \phi^2 F^2} + \underbrace{\frac{c_2^2}{g^4_{SM}} \frac{m_W^2\, E^2}{\Lambda^4}}_{\phi^2 F^2 \times \phi^2 F^2} + \cdots  \Big  ]~,
\end{split}
\end{equation}
where the subscript $L$ refers to the longitudinal mode of the gauge bosons. Eq.~(\ref{eq:TTTL})  implies that the interference term between the SM amplitude and the BSM one with one insertion of the $\phi^2 F^2$ operator is not suppressed compared to the quadratic term. The energy window where the dim-6$\times$dim-6 dominates over the interference is not different from the case without a suppression, that is,
\begin{equation}
 \Lambda \left ( \frac{g_{SM}}{\sqrt{c_2}} \right ) < E < \Lambda~.
 \label{eq:TTTLrange}
\end{equation}  
This is the standard situation,  where the domination of the quadratic term within the EFT validity range arises only for $c_2 \gg g_{SM}$, that is for a strongly coupled UV completion. In the Warsaw basis, also the operators $O_{3W}$ and $O_{H\psi}$ contribute with terms growing with the energy to the $TL$ helicity cross section, and one can show that they lead to the same energy dependence as in  Eq.~(\ref{eq:TTTL}).

\begin{figure}[t]
\centering
\minipage{0.48\textwidth}
  \includegraphics[width=.9\linewidth]{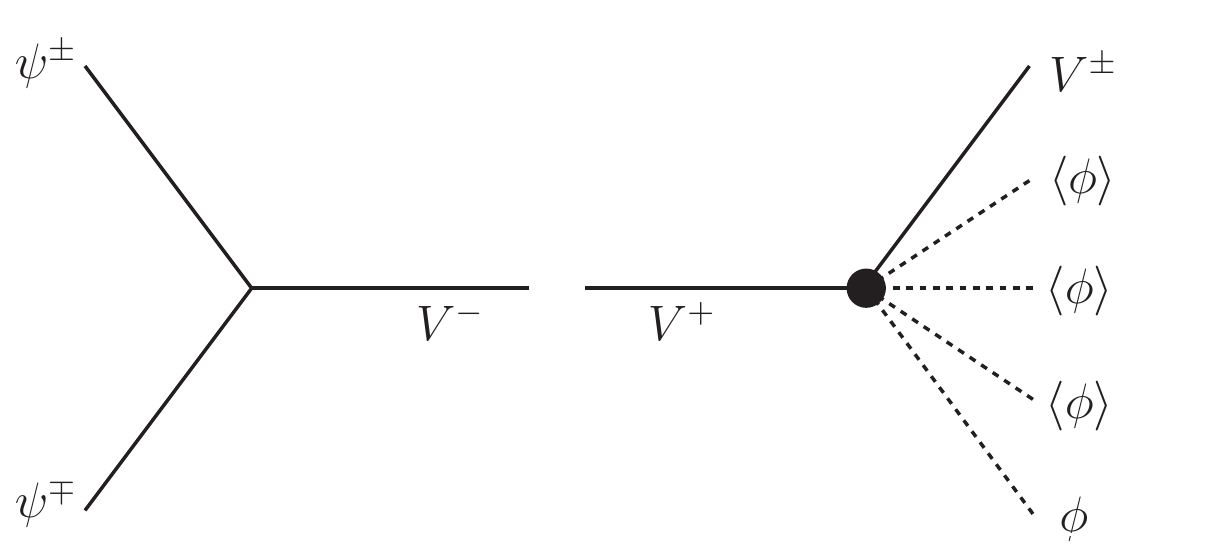}
\endminipage
\minipage{0.48\textwidth}
  \includegraphics[width=.9\linewidth]{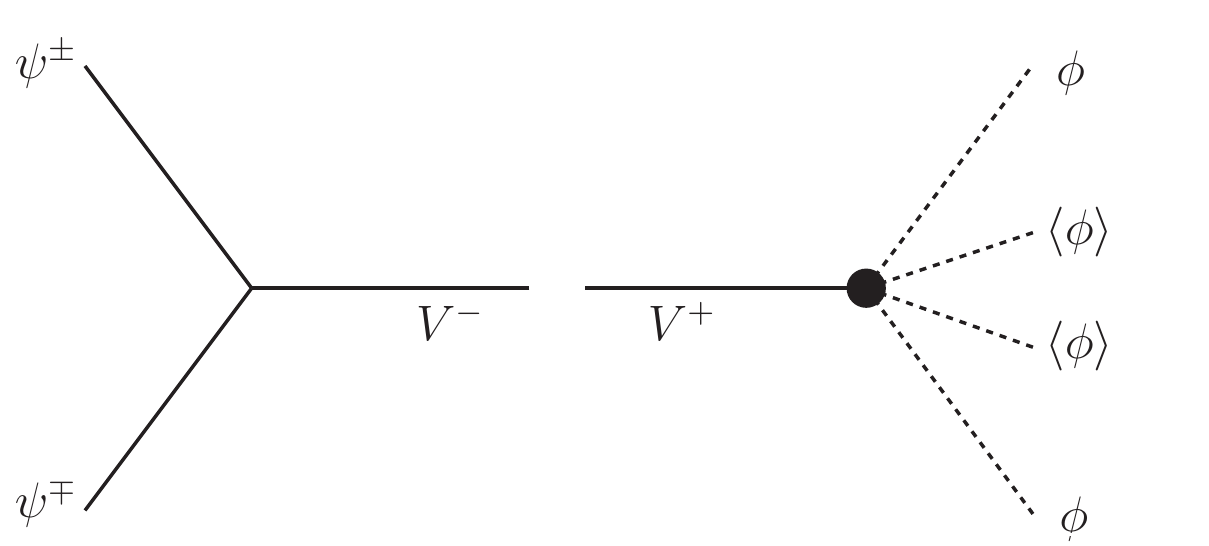}
\endminipage
\caption{The four-point amplitude involving $(\phi D \phi)^2$. The $(\phi D \phi)^2$ operator (shown as a dot) contributes to the six-point amplitude with the helicity of 0 or $\pm 2$ (left). Similarly for five-point amplitude with the helicity of $\pm 1$ (right). We show only diagrams with non-negative helicities.}
\label{fig:four-pt-phi4D2}
\end{figure}

Next, we discuss the $(\phi D \phi)^2$ operator with $(w,\, \bar{w}) = (4,\, 4)$. In the Warsaw basis, one such operator, denoted $O_{HD}$, contributes to diboson production. It can contribute to the triple gauge vertex via the six-point amplitude with helicity of 0 or $\pm 2$ by replacing three $\phi$'s with their VEVs (left diagram in Fig.~\ref{fig:four-pt-phi4D2}). The resulting four-point amplitude will have $h(A_4^{BSM}) = \pm 1$, which requires one helicity flip to interfere with the SM amplitude. The contribution to the cross section $\sigma_{\psi\psi\rightarrow TL}$ is sub-leading, compared to Eq.~(\ref{eq:TTTL}). The other possible contribution is via the five-point amplitude with helicity of $\pm 1$ (right diagram of Fig.~\ref{fig:four-pt-phi4D2}). The resulting four-point amplitude has zero helicity, thus it interferes with the SM one. Similarly, contributions to the cross section $\sigma_{\psi\psi\rightarrow LL}$ do not contain any terms growing with energy. We conclude that the contributions of $(\phi D \phi)^2$  become sub-dominant at high energies compared to those of the other operators. 

\begin{figure}[t]
\centering
\minipage{0.3\textwidth}
  \includegraphics[width=0.9\linewidth]{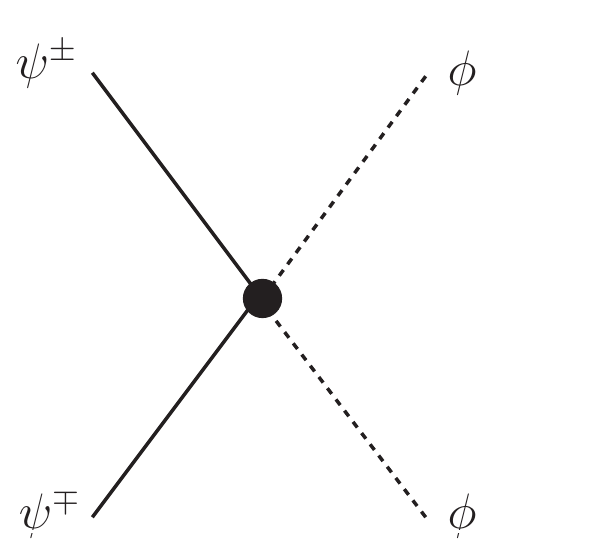}
\endminipage
\caption{The four-point amplitude induced by the contact operator, $\bar{\psi}\gamma\psi\phi D\phi$ with $h (A^{\bar{\psi}\gamma\psi\phi D\phi}_4)=0$.}
\label{fig:four-pt-contact-nosuppression}
\end{figure}
Finally, we consider the operator $\bar{\psi}\gamma\psi \, \phi D\phi$ with $(w,\, \bar{w}) = (4,\, 4)$. The BSM amplitude with the insertion of this operator has the total helicity of zero, and it can interfere with the SM amplitude without any suppression. The contribution to the cross section $\sigma_{\psi\psi\rightarrow LL}$ is thus estimated as 
\begin{equation}\label{eq:contact:TTLL}
\sigma_{\psi\psi \rightarrow LL} = \frac{g^4_{SM}}{E^2} \Big [1 +  \underbrace{\frac{c_4}{g^2_{SM}}\frac{E^2}{\Lambda^2}}_{SM\times \bar{\psi}\gamma\psi\phi D\phi} + \underbrace{\frac{c_4^2}{g^4_{SM}}\frac{E^4}{\Lambda^4}}_{\bar{\psi}\gamma\psi\phi D\phi \times \bar{\psi}\gamma\psi\phi D\phi} + \cdots \Big ]~. 
\end{equation}
In spite of the different energy dependence compared to the $LT$ cross section in \eref{TTTL}, the energy range where the quadratic term dominates over the interference one is analogous as in \eref{TTTLrange}:   
\begin{equation}
 \Lambda \left ( \frac{g_{SM}}{\sqrt{c_4}} \right ) < E < \Lambda~.
\end{equation}
In this case, again, the domination of the quadratic term within the EFT validity range can arise only for  $c_4 \gg g_{SM}$. 
In the Warsaw basis the contributions of other operators than $\bar{\psi}\gamma\psi \, \phi D\phi$ leads to a softer energy dependence. 
The coefficients of these operators are, typically, stringently   constrained by electroweak precision measurements~\cite{Efrati:2015eaa}.   
However, two linear combinations  of these operators with $O_{HD}$ and $O_{HWB}$ do not affect the  the $Z$ and $W$ couplings to fermions, but they do contribute to the aTGCs  $\delta g_{1,z}$ and $\delta \kappa_\gamma$ \cite{Pomarol:2013zra}.\footnote{In the SILH basis these two linear combinations are traded for a combination of purely bosonic operators $O_W$, $O_{B}$, $O_{HW}$ and $O_{HB}$.}
At the LHC, these combinations are probed via diboson production and  Higgs physics.

The remaining contact operators involving two fermions, which could potentially contribute to diboson production,  are the Yukawa-like operators, $\bar{\psi}\psi \phi^3$, and the dipole operators, $F \bar{\psi}\psi \phi$.
However, they both have a L-R (or R-L) chiral structure, which means that they do not interfere with the SM in the limit of massless light quarks. Furthermore, their coefficients are expected to be proportional to light quark Yukawas, providing a further suppression also for the quadratic terms. For these reasons we do not discuss them further.

\section{Helicity Amplitudes for $VV$ production at the LHC}
\label{app:HelAmp}

\begin{figure}[t]
\centering
\minipage{0.45\textwidth}
  \includegraphics[width=.9\linewidth]{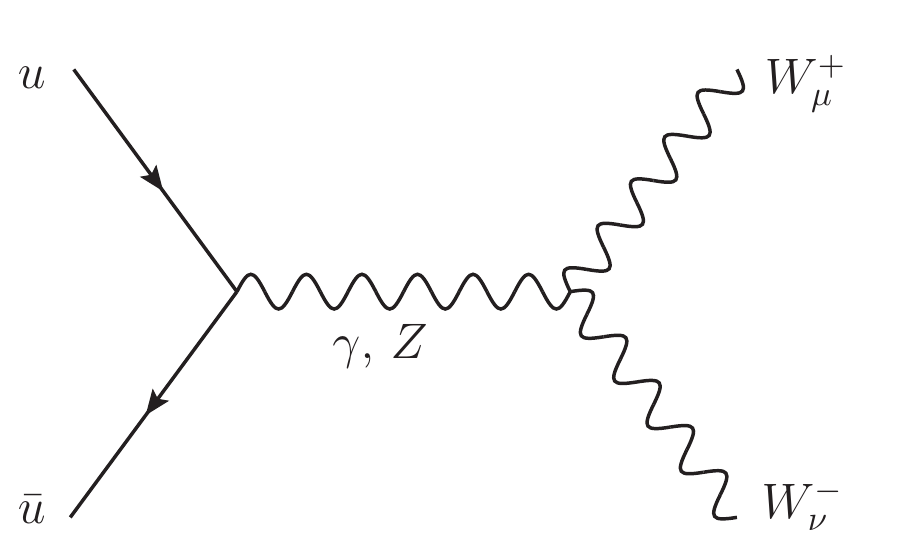}
\endminipage
\quad
\minipage{0.45\textwidth}
 \includegraphics[width=.7\linewidth]{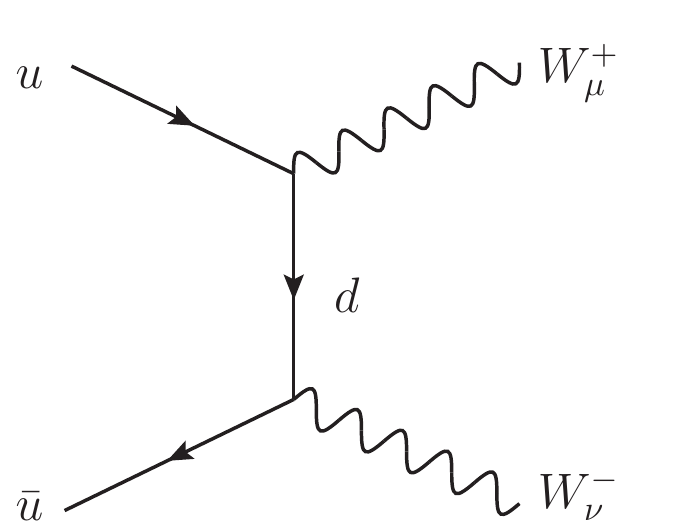}
\endminipage
\caption{The s-channel and t-channel diagrams of $u\bar{u}\rightarrow W^+W^-$. Similar diagrams for down-type initial state quarks.}
\label{fig:aTGC}
\end{figure}

We consider the process $u \bar u \to W^- W^+ $ in the limit of  massless quarks (very similar results hold for  $d \bar d \to W^- W^+$, $u \bar d \to W^+ Z $, and  $\bar u d \to W^- Z $).  
Ref.~\cite{Azatov:2016sqh} pointed out that it is illuminating to expand the helicity amplitudes for this process in $m_W^2/s$, where $\sqrt{s}$ is the center-of-mass energy of the partonic collision.  
In the SM, the amplitudes at the lowest order in $m_W^2/s$ take the particularly simple form: 
\bea
\begin{split}
\cA( -+ \to 00 )&=  {3 g_L^2 + g_Y^2 \over 12} \sin\theta + \cO(m_W^2/s),
\qquad 
\cA( +- \to 00 )=  - {g_Y^2 \over 3}  \sin\theta  + \cO(m_W^2/s), 
\\[0.15cm]
\cA(-+ \to \pm \mp) &=   
-  {\mp 1 +  \cos \theta \over   1 +  \cos \theta} {g_L^2 \over 2} \sin \theta, 
 \qquad  \qquad
\cA( + - \to \pm \mp ) =    0 ,
\\[0.15cm]
\cA(- +  \to \pm 0) & =   \cO(m_W/\sqrt{s}),  \qquad  \qquad \qquad    \cA(+ -  \to  \pm 0 )  =   \cO(m_W/\sqrt{s}), 
\\[0.15cm]
 \cA(- +  \to \pm \pm)  &=     \cO(m_W^2/s),  \qquad  \qquad \qquad     \cA(+ -   \to \pm \pm)  =     \cO(m_W^2/s). 
 \end{split}
 \eea 
 where $g_L$, $g_Y$ are the SM $SU(2) \times U(1)$ couplings, and $\theta$ is the scattering angle of $W^-$.  The amplitudes with  $++$ and $--$ fermion helicities vanish in the limit where the fermions are massless.   

In the presence of aTGCs, the leading correction in $m_W^2/s$ to the helicity  amplitudes are as follows: 
\bea
\begin{split}
\delta  \cA( -+ \to 00 )&= {s \over m_W^2} {g_L^2 \over 12} \sin\theta  \Big [ 
  6 \delta g^{Wq}_L - 6 \delta g^{Zu}_L  - 3 \delta \kappa_z 
-   4 s_\theta^2 (\delta \kappa_\gamma - \delta \kappa_z)   \Big ] + \cO(s^0),
\\[0.15cm]
\delta\cA( +- \to 00 )&= { s \over m_W^2}  {g_L^2 \over 6} \sin\theta \left  [ 
 3 \delta g^{Zu}_R +   2 s_\theta^2  (\delta \kappa_\gamma - \delta \kappa_z)  \right ] +  \cO(s^0), 
\\[0.15cm]
\delta \cA( -+ \to \pm \mp ) &=  \cO(s^0), \qquad  \delta \cA( + - \to \pm \mp ) =    0,   
\\[0.15cm]
\hspace{-1cm}\delta \cA( -+ \to \pm 0 )&=  {\sqrt{s} \over  m_W} {g_L \over 12 \sqrt 2} 
\left (\pm 1-    \cos \theta \right ) \Big  [ 
3 \delta g_{1.z} + 3 \delta \kappa_z  + 3\lambda_z     
\\[0.15cm] 
& \quad +  12  \delta g^{Zu}_L -12 \delta g^{Wq}_L -  4 s_\theta^2  (\delta g_{1,z } - \delta \kappa_\gamma + \delta \kappa_z) \Big ] 
+ \cO(s^{-1/2}),
\\[0.15cm]
\hspace{-1cm} \delta\cA( +- \to  \pm 0 )&=  {\sqrt{s} \over  m_W} {g_L \over 3 \sqrt 2}  \left ( \pm 1 +   \cos \theta \right ) \Big  [ 
 3 \delta g^{Zu}_R -   s_\theta^2   (\delta g_{1,z } - \delta \kappa_\gamma + \delta \kappa_z)  \Big ] +  \cO(s^{-1/2}) , 
\\[0.15cm]
\delta \cA( -+ \to \pm \pm ) &= { s \over m_W^2} {g_L^2 \over 4} \sin\theta \lambda_z  + \cO(s^0), 
\qquad 
\delta \cA( + - \to \pm \pm) = \cO(s^0) .
\end{split}
\eea
Recall that $\delta \kappa_z = \delta g_{1,z} - s_\theta^2 \delta \kappa_\gamma$. 
For completeness, we also display the dependence on the anomalous couplings of $W$ and $Z$ to quarks $\delta g^{Vq}$ (we use the conventions of Ref.~\cite{HiggsBasis}), which also may lead to the growth of the amplitudes with the energy.  
Now, we can see that $\cO(s/m_W^2)$ pieces in the BSM part coincide with the $\cO(s^0)$ piece in the SM  part only for the helicity amplitude with two longitudinal gauge bosons \cite{Azatov:2016sqh}.  
As a result, only the production cross section  of two longitudinal gauge boson will scale with energy in the expected way, that is with $E^{-2}$, $E^0$, and $E^2$  behavior of the SM${}^2$, interference, and BSM${}^2$ terms, respectively. 
For the remaining helicity amplitudes, either the SM or the BSM part carries $m_W/s$ suppression factors, which results in suppressing the interference term compared to naive expectations. 

Using the maps between anomalous couplings and $D=6$ operators in  Ref.~\cite{HiggsBasis}, one can express these results in terms of Wilson coefficients in any of the popular basis. For example, the aTGCs are related to the coefficients in the Warsaw basis by \david{
%
\bea
\label{eq:warsaw_atgc}
\begin{split}
\delta g_{1,z}  &=   - \frac{v^2}{\Lambda^2} {g_L^2  + g_Y^2 \over 4 (g_L^2 - g_Y^2)}  
\left (  4 {g_Y \over g_L}  w_{\phi WB}  +  w_{\phi D}  -    [w_{\ell \ell}]_{1221}  +  2  [  w^{(3)}_{\phi \ell } ]_{11}  +  2  [ w^{(3)}_{\phi\ell } ]_{22}    \right )~, 
\\[0.15cm]
\delta \kappa_\gamma &= \frac{v^2}{\Lambda^2} {g_L \over g_Y}  w_{\phi WB}~, \qquad
\lambda_z  =   - \frac{v^2}{\Lambda^2} {3 \over 2} g_L  w_{W}~, 
\end{split}
\eea}
where we use the original operator normalization of Ref.~\cite{Grzadkowski:2010es} \david{(and \cite{HiggsBasis,deFlorian:2016spz}).}
See also Ref.~\cite{HiggsBasis} for the relation between the vertex correction $\delta g^{Vq}$ and the Wilson coefficients. 
Plugging in these formulas in the helicity amplitudes above, the helicity cross sections schematically take the form,
\bea
\label{eq:helxsec}
\begin{split}
\sigma_{u \bar u \to 00} &\sim  
{g_{\rm SM}^4 \over s } \Big (  1  +  {s \over m_W^2}   \sum_i \alpha_i\, c_{LL}^i   
+  {s^2 \over m_W^4}  \sum_{ij} \alpha_{ij}\, c_{LL}^i c_{LL}^j  \Big )~,
\\[0.15cm]
\sigma_{u \bar u \to \pm 0} & \sim   
{g_{\rm SM}^4  \over s } \Big ( {m_W^2 \over s}  
 +   \sum_i  \beta_i\, c_{LT}^i     +    {s \over m_W^2}  \sum_{ij} \beta_{ij}\, c_{LT}^i c_{LT}^j  \Big )~,
\\[0.15cm]
\sigma_{u \bar u \to \pm \mp} & \sim 
{g_{\rm SM}^4  \over s } \Big ( 1 +   \sum_i  \gamma_i\, c_{TT}^i     +     \sum_{ij} \gamma_{ij}\, c_{TT}^i c_{TT}^j  \Big )~,
\\[0.15cm] 
\sigma_{u \bar u \to \pm \pm} & \sim   
{ g_{\rm SM}^4 \over s } \Big ( {m_W^4 \over s^2 }  
 +   \kappa g_{\rm SM}\,   \bar c_{3W}   + \kappa' {s^2 \over m_W^4} g_{\rm SM}^2\, \bar  c_{3W}^2 \Big )~,
\end{split}
\eea 
where $\alpha,\beta,\gamma,\kappa$'s are numerical $O(1)$ coefficients (in general  depending on $s_\theta$) whose exact values are not relevant for this discussion, and the vectors of Wilson coefficients are defined as 
%
%
\david{\beq
\begin{split}
c_{LL} &=  \frac{v^2}{\Lambda^2}  ( w_{\phi q}^{(3)}, w_{\phi q}^{(1)})~, 
\\[0.15cm] 
c_{LT} &=  \frac{v^2}{\Lambda^2} ( w_{\phi q}^{(3)}, w_{\phi q}^{(1)}, w_{\phi WB}, w_{W})~, 
\\[0.15cm]
c_{TT} &=  \frac{v^2}{\Lambda^2}  ( w_{ \phi q}^{(3)}, w_{\phi q}^{(1)},  w_{\phi WB}, w_{\phi \ell}^{(3)},  w_{\phi D}, [ w_{\ell \ell}]_{1221} )~.
\end{split}
\eeq}
In a similar way we can find how the  SILH basis \cite{Giudice:2007fh} operators affect which helicity amplitude by using the map \cite{HiggsBasis,deFlorian:2016spz}
\begin{equation}
\label{eq:PEL_tgc}
\begin{split}
\delta g_{1z} &=
 - {g_L^2 + g_Y^2 \over g_L^ 2 - g_Y^2} \left [
 {g_L^2 - g_Y^2 \over g_L^2 } \bar c_{HW} + \bar c_W + \bar c_{2W} + {g_Y^2 \over g_L^2} \bar c_B +  {g_Y^2 \over g_L^2} \bar c_{2B}  - {1 \over 2}  \bar c_T  \david{+ \frac{1}{2} [\bar c^\prime_{H\ell}]_{22}} \right ]~,
\\[0.15cm]  
\delta \kappa_\gamma  &=  -  \bar c_{HW}  - \bar c_{HB}~,  \qquad  \lambda_z =   - 6  g_L^2 \bar c_{3W}~,  
\end{split}
\end{equation} 
where we use the notation and normalizations of Ref.~\cite{Contino:2013kra}. 
In the SILH basis the helicity cross sections take the same form as in \eref{helxsec} with $c_{3W} \to g_{\rm SM}c_{3W}$ and
\bea
\begin{split}
c_{LL} &=    (\bar c_{Hq}', \bar c_{Hq},\bar c_{2B}, \bar c_{2W}, \bar c_W, \bar c_B, \bar c_{HB}, \bar c_{HW} )~, 
\\[0.15cm]
c_{LT} &=    (\bar c_{Hq}', \bar c_{Hq},\bar c_{2B}, \bar c_{2W}, \bar c_W, \bar c_B, \bar c_{HB}, \bar c_{HW}, \bar c_{3W})~,
\\[0.15cm]
c_{TT} &=    (\bar c_{Hq}', \bar c_{Hq},\bar c_{2B}, \bar c_{2W}, \bar c_W, \bar c_B,   \bar c_T)~.
\end{split}
\eea


\end{appendix}


\bibliographystyle{JHEP}

\providecommand{\href}[2]{#2}\begingroup\raggedright\endgroup

\end{document}